\documentclass[twocolumn]{aastex63}
\usepackage{amsmath}

\newcommand{\casa}{{\sc casa}}
\newcommand{\sfrsd}{$\Sigma_{\mathrm{SFR}}$}

\newcommand{\jupone}{CO(1$-$0)}
\newcommand{\juptwo}{CO(2$-$1)}
\newcommand{\jupthree}{CO(3$-$2)}
\newcommand{\jupfour}{CO(4$-$3)}
\newcommand{\jupfive}{CO(5$-$4)}

\newcommand{\Htwo}{\mbox{\rm H$_{2}$}}
\newcommand{\CII}{\mbox{\rm [C{\small II}]}}
\newcommand{\CI}{\mbox{\rm [C{\small I}]}}
\newcommand{\OIII}{\mbox{\rm [O{\small III}]}}
\newcommand{\OI}{\mbox{\rm [O{\small I}]}}

\received{}
\revised{}
\accepted{}
\submitjournal{}

\shorttitle{Resolved CO Excitation in DYNAMO}
\shortauthors{Lenki\'{c} et al.}
\graphicspath{{./}{Figures/}}

\begin{document}

\title{CO Excitation in High$-z$ Main Sequence Analogues: Resolved \jupfour/\jupthree{} Line Ratios in DYNAMO Galaxies}

\correspondingauthor{Laura Lenki\'{c}}
\email{laura.lenkic@nasa.gov, llenkic@usra.edu}

\author[0000-0003-4023-8657]{Laura Lenki\'{c}}
\affiliation{Department of Astronomy, University of Maryland, College Park, MD 20742, USA}
\affiliation{SOFIA Science Center, USRA, NASA Ames Research Center, M.S. N232-12, Moffett Field, CA 94035, USA}

\author[0000-0002-5480-5686]{Alberto D. Bolatto}
\affiliation{Department of Astronomy, University of Maryland, College Park, MD 20742, USA}

\author[0000-0003-0645-5260]{Deanne B. Fisher}
\affiliation{Centre for Astrophysics and Supercomputing, Swinburne University of Technology, PO Box 218, Hawthorn, VIC 3122, Australia}
\affiliation{ARC Centre of Excellence for All Sky Astrophysics in 3 Dimensions (ASTRO 3D)}

\author[0000-0002-4542-921X]{Roberto Abraham}
\affiliation{Department of Astronomy \& Astrophysics, University of Toronto, 50 St. George Street, Toronto, ON M5S 3H4,
Canada}

\author[0000-0003-2508-2586]{Karl Glazebrook}
\affiliation{Centre for Astrophysics and Supercomputing, Swinburne University of Technology, PO Box 218, Hawthorn, VIC 3122, Australia}
\affiliation{ARC Centre of Excellence for All Sky Astrophysics in 3 Dimensions (ASTRO 3D)}

\author{Rodrigo Herrera-Camus}
\affiliation{Departamento de Astronomía, Universidad de Concepción, Barrio Universitario, Concepción, Chile}

\author[0000-0003-2508-2586]{Rebecca C. Levy}
\altaffiliation{NSF Astronomy and Astrophysics Postdoctoral Fellow}
\affiliation{Steward Observatory, University of Arizona, Tucson, AZ 85721, USA}

\author[0000-0002-1527-0762]{Danail Obreschkow}
\affiliation{International Centre for Radio Astronomy Research (ICRAR), University of Western Australia, Crawley, WA 6009, Australia}

\author{Carrie Volpert}
\affiliation{Department of Astronomy, University of Maryland, College Park, MD 20742, USA}



\begin{abstract}
The spectral line energy distribution of carbon monoxide contains information about the physical conditions of the star forming molecular hydrogen gas; however, the relation to local radiation field properties is poorly constrained. Using $\sim 1-2$~kpc scale ALMA observations of \jupthree{} and \jupfour{}, we characterize the \jupfour/\jupthree{} line ratios of local analogues of main sequence galaxies at $z \sim 1-2$, drawn from the DYNAMO sample. We measure \jupfour/\jupthree{} across the disk of each galaxy and find a median line ratio of $R_{43} = 0.54${\raisebox{0.5ex}{\tiny$^{+0.16}_{-0.15}$}} for the sample. This is higher than literature estimates of local star-forming galaxies and is consistent with multiple lines of evidence that indicate DYNAMO galaxies, despite residing in the local Universe, resemble main-sequence galaxies at $z \sim 1-2$. Comparing to existing lower resolution \jupone{} observations, we find $R_{41}$ and $R_{31}$ values in the range $\sim 0.2-0.3$ and $\sim 0.4-0.8$ respectively. We combine our kpc-scale resolved line ratio measurements with HST observations of H$\alpha$ to investigate the relation to star formation rate surface density and compare this relation to expectations from models. We find increasing \jupfour/\jupthree{} with increasing star formation rate surface density; however, models over-predict the line ratios across the range of star formation rate surface densities we probe, particularly at the lower range. Finally, SOFIA observations with HAWC+ and FIFI-LS reveal low dust temperatures and no deficit of \CII{} emission with respect to the total infrared luminosity. 
\end{abstract}



\section{Introduction} \label{sec:intro}
Molecular hydrogen gas (\Htwo{}) is routinely mapped in high-redshift (high$-z$) galaxies with instruments such as the Atacama Large Millimeter/sub-millimeter Array (ALMA) and NOrthern Extended Millimeter Array (NOEMA) through the use of high rotational lines (high$-J$) of carbon monoxide \citep[CO; see e.g.,][]{genzel10,tacconi10,decarli14,walter16,freundlich19}. High$-J$ lines of CO can be used to address important topics such as the evolution of molecular gas reservoirs in galaxies across cosmic time \citep[see e.g.,][]{walter14,decarli16,riechers19,decarli19,lenkic20}. Mapping \Htwo{} through high$-J$ CO emission in high$-z$ galaxies provides certain advantages over \jupone{}, because these lines are bright and allow for higher resolution observations. However, limited constraints of the CO excitation ladder, or the CO spectral line energy distribution (SLED), render the conversion to the ground state transition, \jupone{}, uncertain. The CO excitation ladder contains information about the temperature and density of the \Htwo{} material \citep[see e.g.,][for a review]{carilli13}, and understanding how those properties relate to local star formation activity will improve our understanding of how high$-J$ CO lines map to \jupone{} and \Htwo{} mass.

Several studies have characterized the CO excitation ladder in various local galaxy populations. In the Milky Way galactic center, observations from the COBE Far Infrared Absolute Spectrophotometer, which constrain the CO SLED from $J=1-0$ to $J=8-7$, show that the line ratios can be modeled with an excitation temperature of 40~K and that the CO SLED peaks at $J=3$ \citep{fixsen99}. A recent systematic study of \jupone{} to \jupthree{} in nearby galaxies finds that the Rayleigh-Jeans brightness temperature ratios are generally higher in galaxy centers, decreasing with radius and diminishing star formation rate (SFR) surface density \citep{leroy22}.

\citet{kamenetzky16} study CO emission up to $J = 13-12$ in ultraluminous infrared galaxies \citep[ULIRGS; see also][]{greve14}, active galactic nuclei (AGN), and non-ULIRGs to find that CO SLEDs peak at increasingly higher$-J$ with increasing far-infrared (FIR) luminosity, indicating higher kinetic temperatures or densities are required \citep[see also Figure 1 of][for how CO excitation depends on galaxy type and excitation temperature]{obreschkow09}. Observations of submillimeter galaxies (SMGs) show that they also have CO SLEDs with an ``excess'' of CO excitation with respect to the Milky Way; these generally rise up to $J=5$ and then turn over for higher rotational states \citep{bothwell13, spilker14}. These high excitation CO SLEDs suggest that alternative heating sources are required in these extreme galaxies such as mechanical heating via shocks, turbulence, or cosmic rays.

Although several large studies probe the CO SLEDs of extreme systems like SMGs and U/LIRGS, the CO SLEDs of normal $z\sim1-2$ star-forming galaxies are not as well characterized. \citet{valentino20} conduct a large survey of mid$-$ and high$-J$ CO lines with ALMA in main sequence galaxies at $z \sim 1.1 - 1.7$, and find that they have higher excitation than the Milky Way but are not quite as highly excited as ULIRGs, SMGs, or QSOs. \citet{daddi15} find similar results in a sample of four main-sequence near-IR selected galaxies at $z \sim 1.5$, using \juptwo{}, \jupthree{}, and \jupfive{} observations. \citet{bolatto15} also present the Rayleigh-Jeans brightness temperature \jupthree{}/\jupone{} line ratio of four main sequence galaxies observed with the Plateau de Bure Interferometer (PdBI), and find a ratio of about unity denoting high excitation. Finally, multiple case studies of the CO ladder in specific galaxies exist \citep[see e.g.,][]{barvainis97,vanDerWerf10,kamenetzky12,aravena14,dessauges-zavadsky17,brisbin19,sharon19,henriquez-brocal22,klitsch22}, and show that while general trends exist in different galaxy populations, the CO ladder of every galaxy is unique. 

This indicates that it is necessary to understand how CO emission and excitation vary within galaxies and how they relate to other physical properties in order to correctly interpret \Htwo{} masses derived from high$-J$ transitions. This requires resolved studies of CO line ratios; however, observational limitations at high$-z$ make this challenging. To address this, we present a sample of nine galaxies drawn from the DYnamics of Newly Assembled Massive Objects \citep[DYNAMO;][]{green14} observed by ALMA in \jupthree{} and \jupfour{} on $\sim 1-2$~kpc scales. DYNAMO galaxies are nearby ($z \sim 0.1$) objects with high gas fractions, high star formation rates, and widespread turbulence, consistent with known properties of high$-z$ main-sequence galaxies, and many are indeed found to lie on the main-sequence of star formation at $z \sim 1-2$ \citep{fisher19}. Their resemblance to high$-z$ systems and proximity allows us to probe the CO excitation in gas-rich, turbulent galaxies at scales that are not yet achievable at $z \sim 2$ in unlensed systems. Furthermore, theories seek to explain CO line ratios by their local radiation field properties \citep{lagos12,narayanan14,popping14,bournaud15}, and our ALMA observations allow us to compare to model expectations. 

This paper is structured as follows: \S\ref{sec:obs} describes our observations, data reduction, and methods; \S\ref{sec:results} and \S\ref{sec:discussion} describe and discuss our results, and finally we conclude in \S\ref{sec:conclusion}. Throughout this work, we assume $\Lambda$CDM cosmology with H$_{0} = 69.6$~km\,s$^{-1}$, $\Omega_{m} = 0.286$, and $\Omega_{\Lambda} = 0.714$ \citep{bennett14}, and a Kroupa initial mass function \citep[IMF;][]{kroupa01}.

\section{Observations and Data Reduction} \label{sec:obs}
The DYNAMO sample of galaxies was first defined by \citet{green14}, who selected galaxies from the MPA-JHU Value Added Catalog of the Sloan Digital Sky Survey DR4 \citep[SDSS;][]{adelman-mccarthy16} based on their redshift and H$\alpha$ emission. The sample consists of 67 galaxies, half of which have L$_{\mathrm{H}\alpha} > 10^{42}$~erg\,s$^{-1}$ in the 3\arcsec\ diameter SDSS fiber, lying in two redshift windows centered at $z \sim 0.075$ and $z \sim 0.13$. Their stellar masses range from $10^{9}-10^{11}$~M$_{\odot}$ and their SFRs from $\sim 0.1-100$~M$_{\odot}$\,yr$^{-1}$, while their metallicities are about solar \citep{tremonti04}. Employing integral-field spectroscopy of H$\alpha$, \citet{green14} derive H$\alpha$ rotation curves and find high ionized gas velocity dispersions with a mean of $\sim 50$~km\,s$^{-1}$, and gas fractions as high as $f_{gas} \sim 0.8$ ($f_{gas} \equiv M_{gas}/(M_{gas}+M_{*})$; $M_{gas} = M_{HI} + M_{H_{2}}$). Furthermore, they find that DYNAMO galaxies are more ``turbulent'' than local disks, as parameterized by their ratio of rotation velocity to velocity dispersion (V/$\sigma$). These properties make DYNAMO galaxies promising candidates for local analogues of high-redshift, star-forming galaxies. 

Here, we consider a sub-set of nine DYNAMO galaxies that have robustly been identified as consistently more similar to $z \sim 1-2$ star-forming systems: DYNAMO C13-1, C22-2, D13-5, D15-3, G04-1, G08-5, G14-1, G20-2, and SDSS J013527.10-103938.6 (hereafter SDSS 013527-1039). This builds on a multi-wavelength campaign to investigate the nature of star formation at high redshift \citep{bassett14,fisher14,obreschkow15,bassett17,fisher17a,fisher17b,white17,girard21,lenkic21,ambachew22,white22}. All galaxies in our sample are classified as rotating disks based on their H$\alpha$ kinematics. Galaxies C22-2, G04-1, G14-1, and G20-2 are furthermore classified as ``compact'' rotating disks, because their SDSS $r-$band exponential scale lengths are smaller than 3~kpc. For these, the poorer resolution results in less reliable kinematic classifications \citep{green14}. 

All galaxies in our sample have \jupone{} observations from either the PdBI or NOEMA, from which molecular gas fractions ($M_{H_{2}}/(M_{H_{2}}+M_{*})$) of $f_{gas} \sim 20-30$~\% and molecular gas depletion times of $t_{dep} \sim 0.5$~Gyr are inferred \citep{fisher14,white17,fisher19}. These high molecular gas fractions are consistent with those of $z \sim 1-2$ main-sequence star-forming galaxies \citep{daddi10,tacconi10,tacconi13,genzel15,tacconi20}. Similarly, subsequent studies of the gas kinematics in these galaxies consistently show that they do indeed have high ionized gas velocity dispersions \citep{bassett14,oliva-altamirano18,girard21}, similar to main-sequence galaxies at $z \sim 1-2$ \citep{forsterschreiber06,ubler19}. In addition, \citet{fisher17b,white17} both show that these DYNAMO galaxies are consistent with marginally stable disks (Toomre $Q \sim 1$). DYNAMO galaxies also conform to established definitions of clumpy galaxies \citep{fisher17b} at high-redshift \citep[e.g., CANDELS;][]{guo15}. Finally, their clumps are arranged within their host disks such that the redder clumps are preferentially more centrally located than the bluer ones \citep{lenkic21,white22}, which has also been observed in $z \sim 1-2$ clumpy galaxies \citep{forsterschreiber11,soto17,guo18}.

\subsection{ALMA CO Observations} \label{subsec:obs_alma}
We make use of the \jupthree{} and \jupfour{} observations of nine DYNAMO galaxies with the Atacama Large Millimeter/Submillimeter Array (ALMA), associated with project code 2017.1.00239.S (PI: D. B. Fisher). Observations were taken in Band 7 ($275-373$~GHz) and Band 8 ($385-500$~GHz) between 2018-06-01 and 2018-07-10. The spectral windows were configured with bandwidths of 2.00 GHz and channel widths of 15.625 MHz (128 channels). In addition, we also make use of higher resolution \jupthree{} ALMA observations of three DYNAMO galaxies (G04-1, G08-5, and G14-1) associated with the project code 2019.1.00447.S (PI: R. Herrera-Camus). These observations were taken in Band 7 between 2019-10-09 and 2019-10-10. The spectral windows were configured with bandwidths of 1.875 GHz and channel widths of 7.8125 MHz (240 channels). The data associated with both projects were presented in \citet{girard21}.

\begin{figure*}
    \centering
    \includegraphics[width=\textwidth]{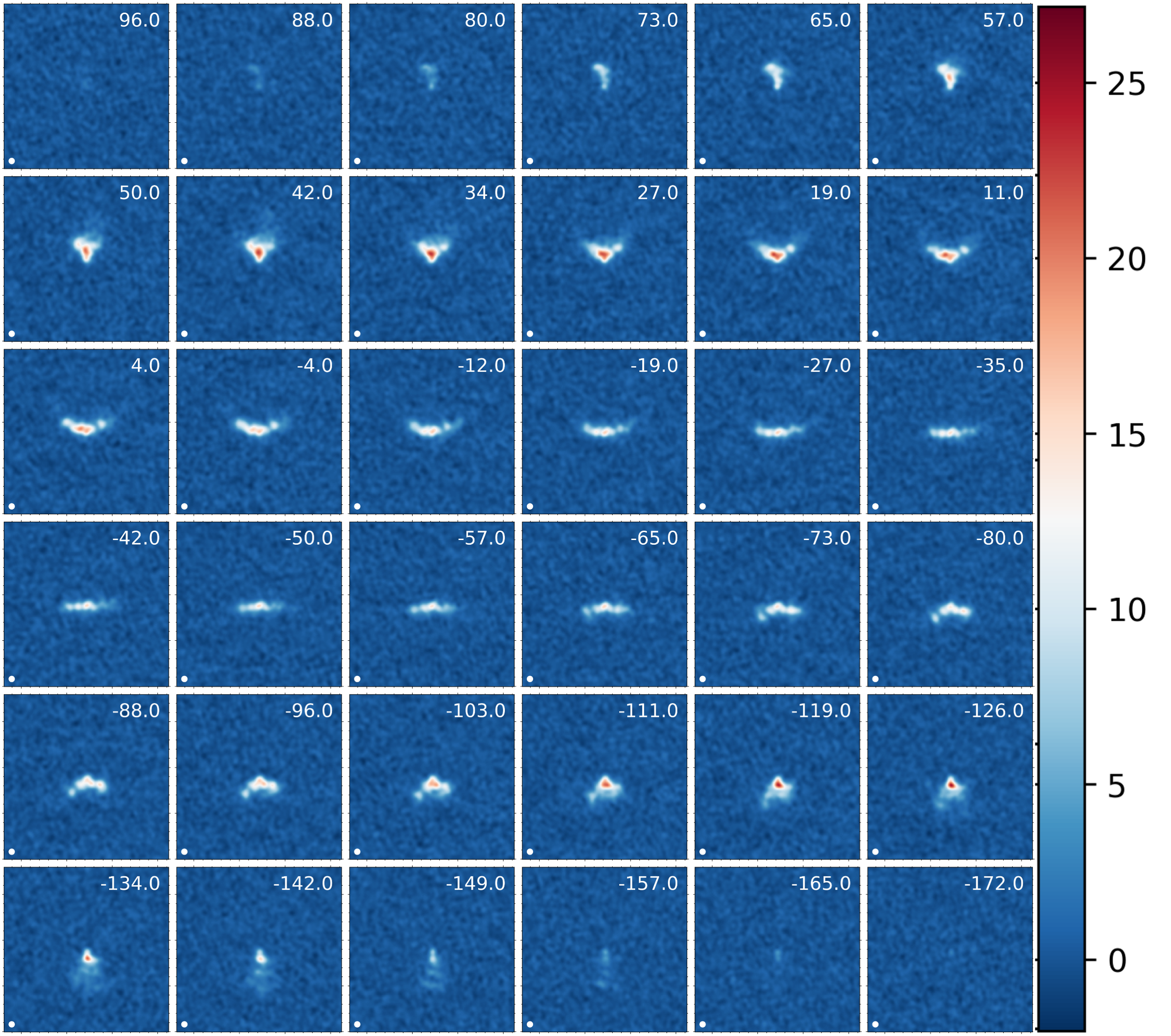}
    \caption{Channel maps of CO(3$-$2) in brightness units of Jy\,beam$^{-1}$ for the galaxy DYNAMO G04-1. Each panel is centered at 04h12m19.713s, -05d54m48.62s, and is 10.8$\times$10.8\arcsec\ in size. The velocity range is $-$172 to 96~km\,s$^{-1}$ in steps of $\sim$8~km\,s$^{-1}$, as indicated in the top right corners. The circularized beam is shown in white in the bottom left corner of each panel. The complete figure set (14 images) is available in the online journal.}
    \label{fig:chann_maps}
\end{figure*}

The visibilities were calibrated and flagged by the observatory with the Common Astronomy Software Application \citep[\casa,][]{mcmullin07} pipeline versions listed in the fifth column of Table \ref{tab:obs}. After calibrating the visibilities, we imaged each observation using \texttt{tclean} in \casa\ version 6.1.0.188 with parameters \texttt{deconvolver=`hogbom'}, \texttt{weighting=`briggs'}, \texttt{robust=0.5}, \texttt{usemask=`auto-multithresh'}, and \texttt{restfreq} set to the redshifted frequency of the observed CO line. We cleaned the data until the residuals were consistent with the root-mean-square (rms) noise levels that are listed in the fourth column of Table \ref{tab:obs}. To derive these thresholds, we consider data cubes with just a shallow clean, mask the emission (see below), and calculate the standard deviation of the masked cubes; i.e., non-line channels. These values are listed in column four of Table \ref{tab:obs}, and we re-clean the data cubes to that rms level. For visualization purposes and ease of comparison to the H$\alpha$ maps, we convolve the final cubes to a circular beam, listed in the second (angular size) and third (physical size) columns of Table \ref{tab:obs}, with the \casa\ \texttt{imsmooth} function. At the redshifts of DYNAMO galaxies in our sample, the beam sizes correspond to physical scales of $\sim 1-2$~kpc. Finally, we export all data cubes with the spectral axis in units of velocity, in the local standard of rest frame, adopting the radio convention. We present channel maps of \jupthree{} for DYNAMO G04-1 in Figure \ref{fig:chann_maps} to show an example of the final data, with the circularized beam shown in the bottom left corner of each panel. The complete figure set (14 images) is available in the online journal.

\begin{deluxetable}{lcccc}
\tablecaption{CO Data Cube Parameters} 
\label{tab:obs}
\tablewidth{700pt}
\tabletypesize{\footnotesize}
\tablehead{
	\colhead{CO Trans.}             &
	\multicolumn{2}{c}{Beam FWHM}   &
	\colhead{rms Noise}             &
	\colhead{{\sc casa} Cal.}       \\
	\colhead{}                      &
	\colhead{(arcsec)}              &
	\colhead{(kpc)}                 &
	\colhead{(mK)}                  &
	\colhead{}                             
} 
\startdata
\sidehead{DYNAMO C13-1}
$3-2$ & 1.07 & 1.60 & 11.5 & v5.1.1-5 \\
\hline
\sidehead{DYNAMO C22-2}
$3-2$ & 1.07 & 1.46 & 16.0 & v5.1.1-5 \\
$4-3$ & 0.81 & 1.11 & 12.5 & v5.1.1-5 \\
\hline
\sidehead{DYNAMO D13-5}
$3-2$ & 1.10 & 1.58 & 8.2 & v5.1.1-5 \\
$4-3$ & 0.79 & 1.14 & 12.1 & v5.1.1-5 \\
\hline
\sidehead{DYNAMO D15-3}
$4-3$ & 0.96 & 1.24 & 10.8 & v5.1.1-5 \\
\hline
\sidehead{DYNAMO G04-1}
$3-2$ & 0.42 & 0.98 & 29.1 & v5.6.1-8 \\
$4-3$ & 0.84 & 1.96 & 21.5 & v5.1.1-5 \\
\hline
\sidehead{DYNAMO G08-5}
$3-2$ & 0.40 & 0.95 & 32.4 & v5.6.1-8 \\
\hline
\sidehead{DYNAMO G14-1}
$3-2$ & 0.43 & 1.02 & 3.3 & v5.6.1-8 \\
$4-3$ & 0.85 & 2.01 & 6.3 & v5.1.1-5 \\
\hline
\sidehead{DYNAMO G20-2}
$3-2$ & 1.23 & 3.08 & 3.7 & v5.1.1-5 \\
$4-3$ & 0.86 & 2.15 & 4.8 & v5.1.1-5 \\
\hline
\sidehead{SDSS 013527-1039}
$3-2$ & 1.23 & 2.81 & 5.2 & v5.1.1-5 \\
$4-3$ & 0.86 & 1.97 & 4.5 & v5.1.1-5 \\
\enddata
\end{deluxetable}

We produce moment zero maps (integrated intensity) by first masking each cleaned data cube along both the spatial and spectral axes. To produce our masks, we first smooth each cleaned data cube to twice the circularized beam full width at half maximum (FWHM). We then compute the rms of the data cube, and mask all pixels that are below $3\times$ the cube rms. For the remaining pixels, we compute the integrated intensity over the channels that are not masked out. We do this for both the \jupthree{} and \jupfour{} observations. Figure \ref{fig:sum_obs} presents these moment zero maps in the two right-most panels.

Finally, for the goal of calculating \jupfour{}/\jupthree{} line ratios, we match the pixel scale and resolution of all 2017 \jupfour{} observations to the pixel scale and resolution of the 2017 \jupthree{} observations, where data for both transitions are available. Similarly, we match the pixel scale and resolution of the 2019 \jupthree{} observations, where available, to the 2017 \jupfour{} observations. We match the pixel scales using the \casa\ function \texttt{imregrid}, while to match the resolution, we use the \casa{} \texttt{imsmooth} tool to convolve the higher resolution data with a Gaussian kernel to produce the lower resolution Gaussian beam. We note that these transformations are done on the cleaned data cubes with the original, non-circular beams, to ensure we are not introducing errors or artifacts in the data. Finally, we apply the masking of the \jupthree{} observations to the \jupfour{} to produce matching integrated intensity maps. This ensures that the intensities we derive for both lines are integrated over the same velocity ranges and regions.

\begin{figure*}
    \centering
    \includegraphics[width=\textwidth]{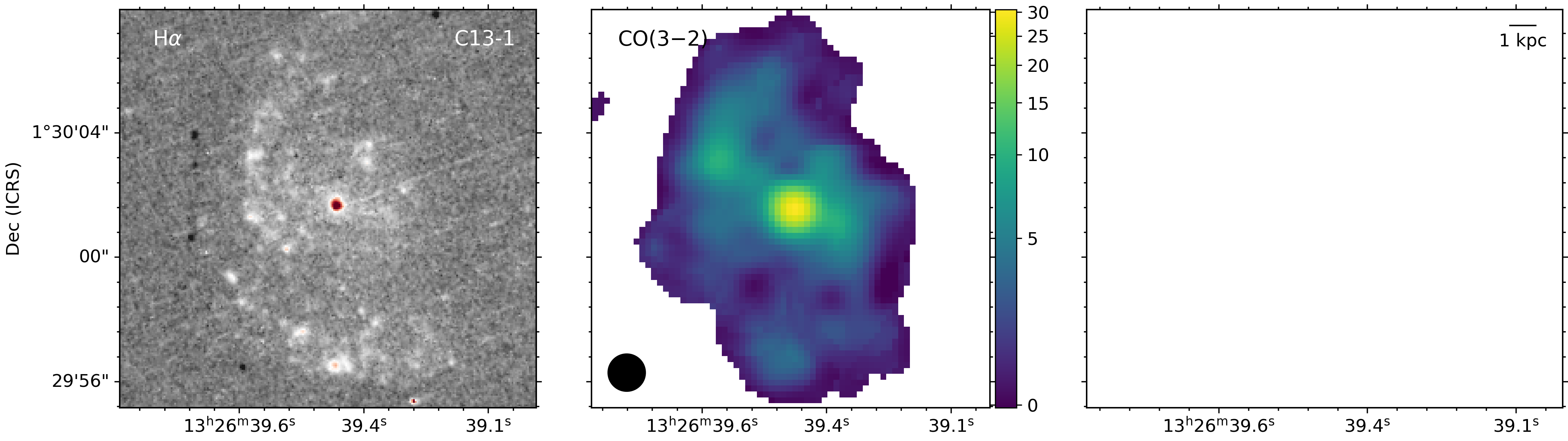}
    
    \includegraphics[width=\textwidth]{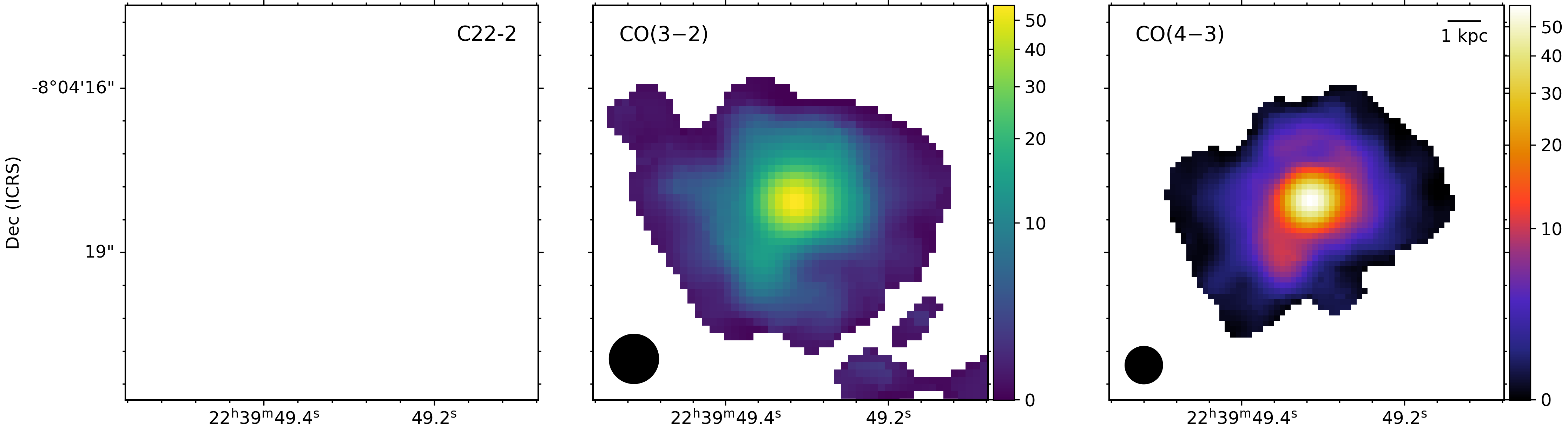}
    
    \includegraphics[width=\textwidth]{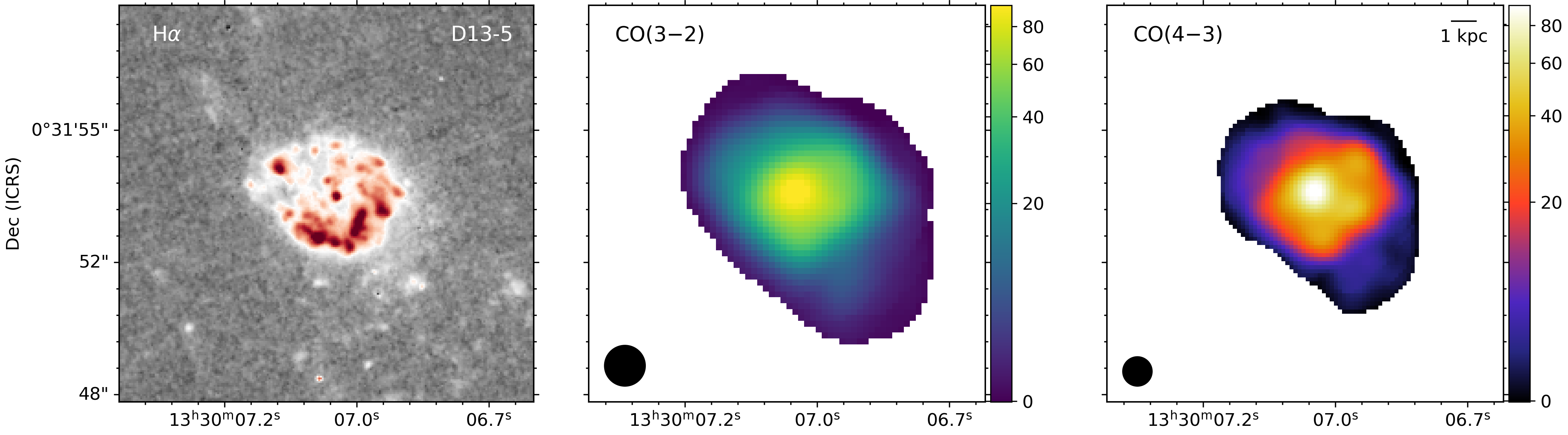}
    
    \includegraphics[width=\textwidth]{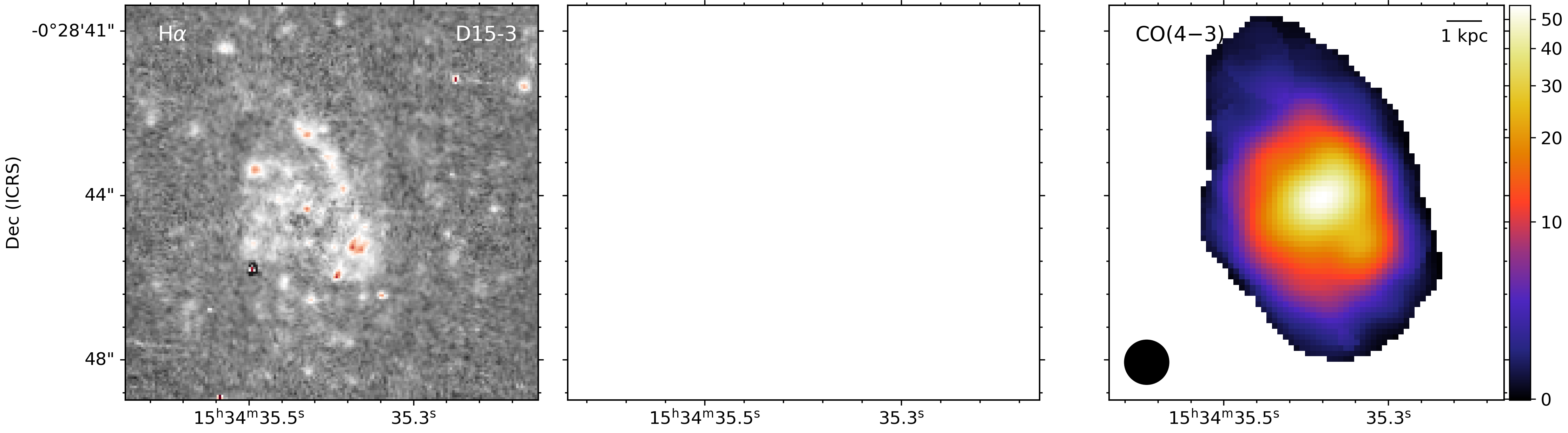}
    
    \includegraphics[width=\textwidth]{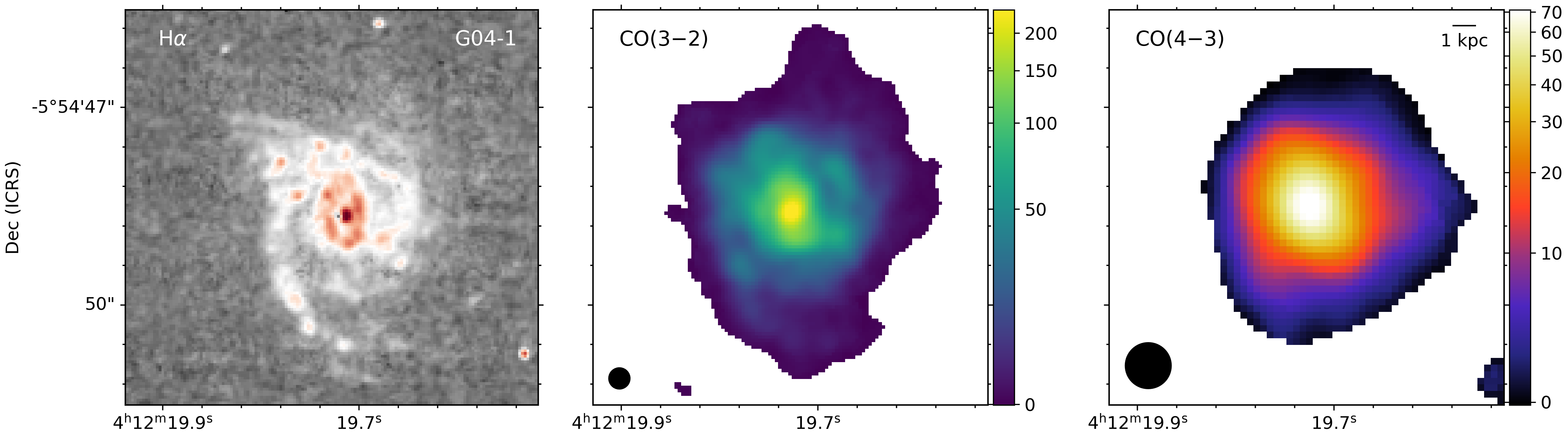}
\end{figure*}

\begin{figure*}
    \centering
    \includegraphics[width=\textwidth]{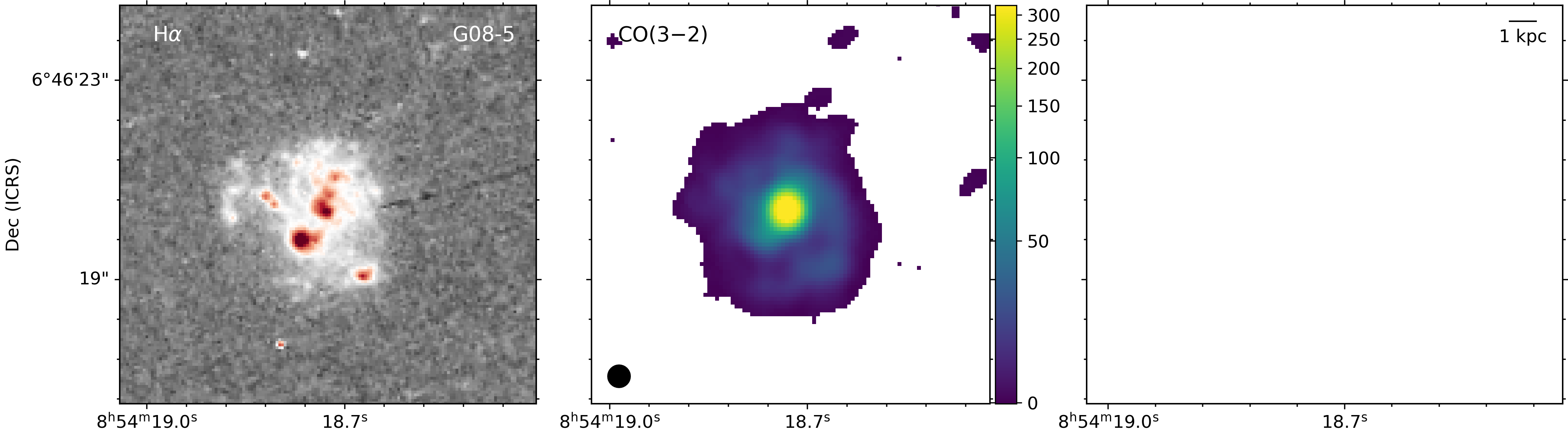}
    
    \includegraphics[width=\textwidth]{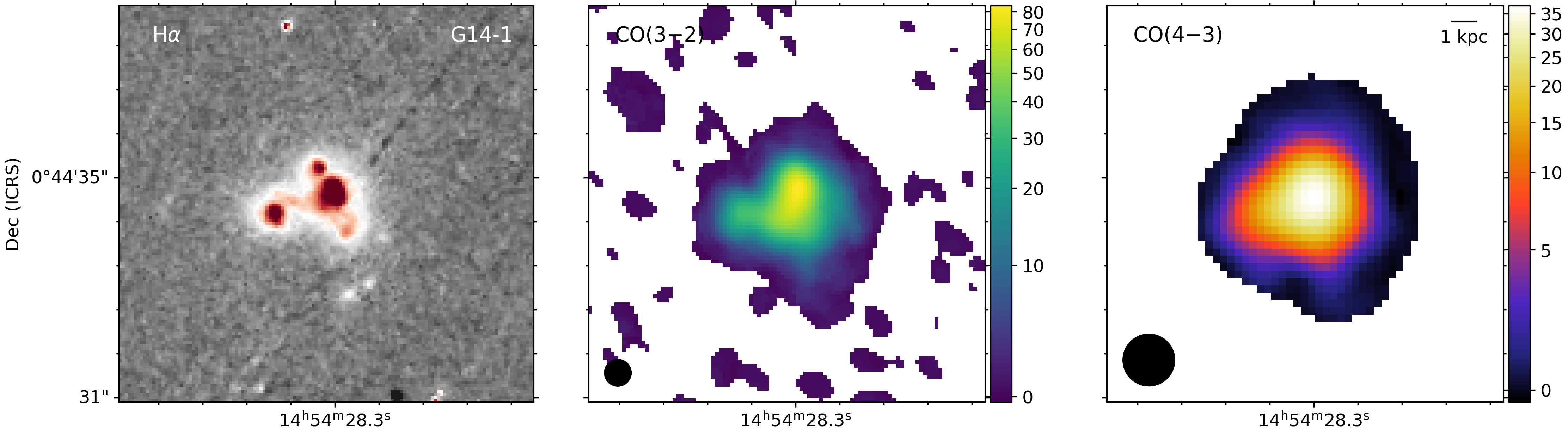}
    
    \includegraphics[width=\textwidth]{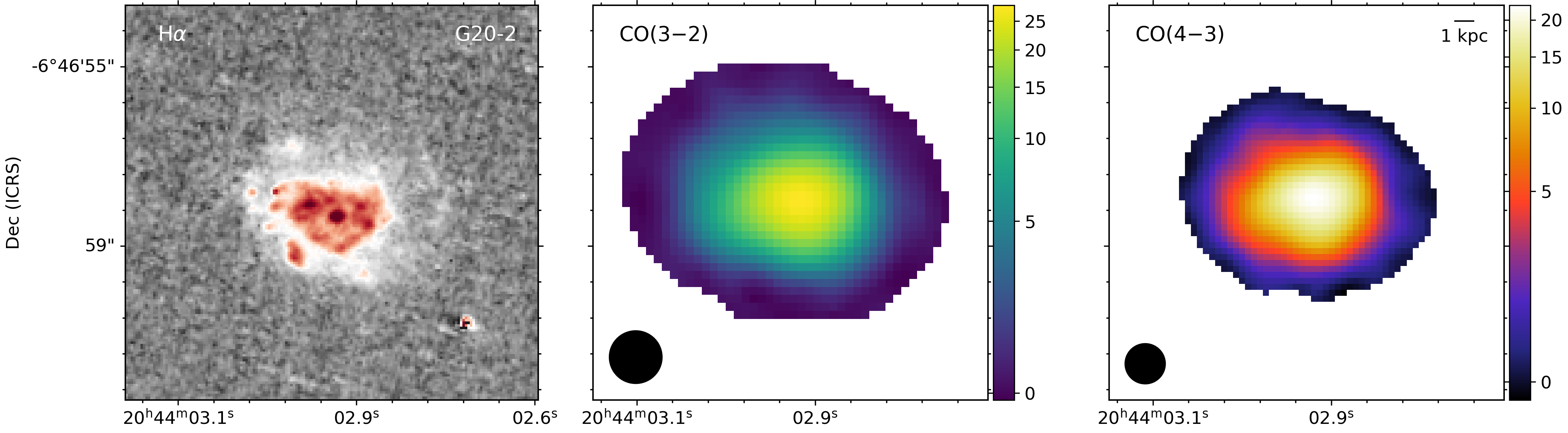}
    
    \includegraphics[width=\textwidth]{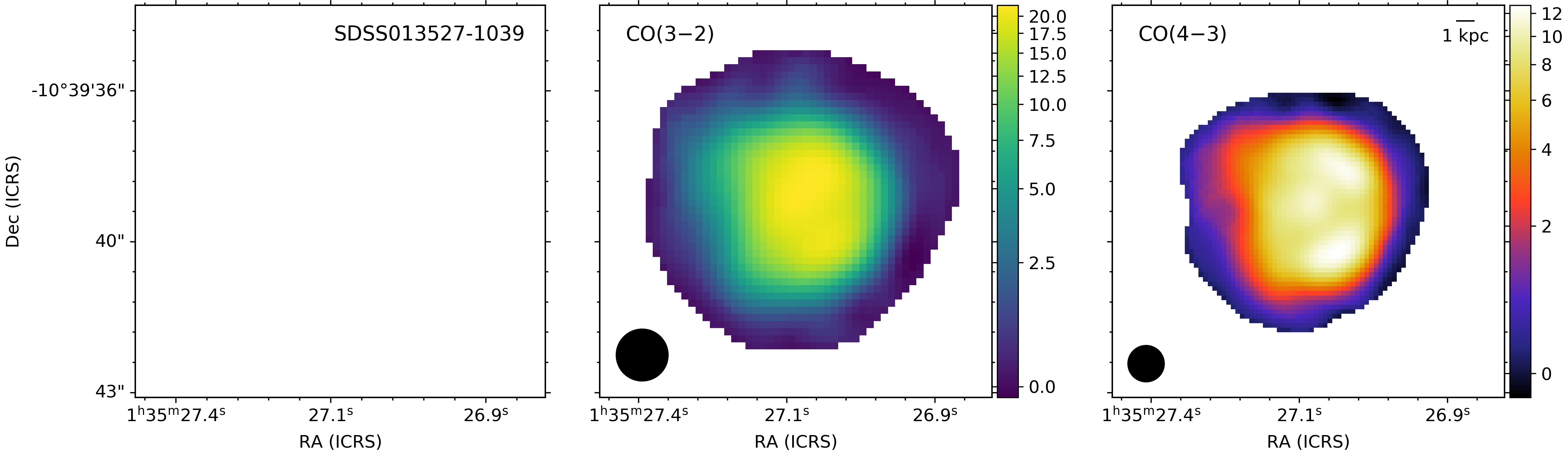}
    \caption{Summary of data sets analyzed in this work for each galaxy in our sample, as indicated in the top right corners of the left-most panels: HST H$\alpha$ (left), \jupthree{} integrated intensity (middle), \jupfour{} integrated intensity (right; both in units of K\,km\,s$^{-1}$). We show all images using an arcsinh stretch. The ALMA CO beam sizes are in the bottom left corners of the middle and right-most panels, while 1~kpc scalebars are shown in the top right corner of the right-most panels. Empty panels indicate that data is absent for the given galaxy.}
    \label{fig:sum_obs}
\end{figure*}

\subsection{HST H$\alpha$ Observations} \label{subsec:obs_hst}
In addition to the ALMA observations of CO, we make use of \textit{Hubble Space Telescope} (HST) observations of H$\alpha$ (PID 12977; P.I.: I. Damjanov) as a tracer of the star formation rate (left-most panel of Figure \ref{fig:sum_obs}). Observations were taken with the Wide Field Camera on the Advanced Camera for Surveys (WFC/ACS) using the FR716N and FR728N narrow-band filters, and were processed with the standard HST pipeline. Continuum observations with the FR647M filter were also taken and used to create continuum-subtracted H$\alpha$ maps \citep[for details, see \S 3.2 of][]{fisher17a}. The final H$\alpha$ maps have a pixel scale of $\sim 0.05$\arcsec\ and a resolution corresponding to physical scales of $\sim 50-200$~pc \citep{fisher17a}.

Our ability to make resolved measurements in these DYNAMO galaxies is limited by the resolution of the ALMA data; therefore, we match the pixel scale and resolution of the H$\alpha$ observations to that of the \jupthree{}, where available, and \jupfour{} otherwise. To achieve this, we convolve each H$\alpha$ observation with a two-dimensional Gaussian function whose FWHM is equal to the circularized beam of the corresponding ALMA observation. Then, we re-project and re-grid the H$\alpha$ observations to match the WCS information and pixel scale of the CO observations using the {\sc Python astropy} package \texttt{reproject}\footnote{\href{https://reproject.readthedocs.io/en/stable/index.html}{https://reproject.readthedocs.io/en/stable/index.html}}, noting that the reproject functions assume that input images have surface brightness units. 

\subsection{SOFIA FIFI-LS and HAWC+ Observations} \label{subsec:sofia_obs}
Finally, we make use of observations from the Stratospheric Observatory for Infrared Astronomy (SOFIA) of DYNAMO galaxies taken by the FIFI-LS \citep{colditz18,fischer18} and HAWC+ \citep{harper18} instruments (PLAN ID 08\_0238 and 09\_158; P.I.: L. Lenki\'{c}) as part of Cycles 8 and 9. The Field-Imaging Far-Infrared Line Spectrometer (FIFI-LS) is an integral field spectrometer with two channels observing simultaneously from $50-125$~\micron{} (blue channel) and $105-200$~\micron{} (red channel). The FIFI-LS observations targeted the \CII{} 158~\micron{} fine-structure emission line in the red channel and the \OIII{} 88~\micron{} fine-structure line (or \OI{} at 63~\micron{} depending on atmospheric transmission) in the blue channel for six galaxies (DYNAMO B08-3, D10-4, D14-1, D15-3, F08-2, F09-1, and F12-4) at 15.6\arcsec{} resolution. These data cover a $1 \times 1$~arcmin$^{2}$ field-of-view (FOV) in the red channel and a $30 \times 30$~arcsec$^{2}$ FOV in the blue channel. FIFI-LS observations were taken on six nights in April 2021 in the nod-match-chop mode, and were reduced using the FIFI-LS pipeline\footnote{\href{https://www.sofia.usra.edu/sites/default/files/2022-09/fifi_users_revL.pdf}{FIFI-LS Redux User's Manual}} \citep{vacca20}. The data reduction steps include ramp fitting and flagging bad pixels, subtracting the chops, wavelength and spatial calibration, flat-field correction, atmospheric transmission correction using the ATRAN models \citep{lord92}, flux calibration, and finally resampling to a regular grid to produce the final data cubes. The observations resulted in \CII{} detections for all galaxies in the sample, and an \OIII{} detection in DYNAMO F08-2 (see Figure \ref{fig:fifi_obs} in Appendix \ref{app:sofia}). 

The High-resolution Airborne Wideband Camera Plus (HAWC+) instrument is a FIR camera and imaging polarimeter with a wavelength coverage of $50-240$~\micron{}. The HAWC+ observations targeted four galaxies (DYNAMO D14-1, D15-3, F08-2, and F12-4) in bands C, D, and E. These data provide measurements of the 89, 155, and 216~\micron{} fluxes at a resolution of 7.8\arcsec{}, 14\arcsec{}, and 19\arcsec{} respectively. Observations were taken on three nights in May 2021 and one night in November 2021 in the on-the-fly mapping mode with a Lissajous scan pattern, and were reduced using the HAWC+ pipeline\footnote{\href{https://www.sofia.usra.edu/sites/default/files/2022-09/hawc_users_revK.pdf}{HAWC+ DRP User's Manual}}. The observations resulted in detections for all galaxies in the sample (see Figure \ref{fig:hawc_obs} in Appendix \ref{app:sofia}).

The typical sizes of galaxies in this sample are $\sim 4$\arcsec{} and our sources are thus point sources for both the FIFI-LS and HAWC+ observations. Appendix \ref{app:sofia} presents the HAWC+ observations in Figure \ref{fig:hawc_obs} and the FIFI-LS integrated intensity maps in Figure \ref{fig:fifi_obs}. While DYNAMO D13-5 is the only galaxy that overlaps with our ALMA sample, we make use of all SOFIA observations described here to measure the SEDs of DYNAMO galaxies, and place the measured dust temperatures within the global context of the line ratio measurements we will present. We also make use of these observations to measure the \CII{} luminosity and measure the \CII{}-to-total far-infrared luminosity ratios.

\subsection{Resolved Measurements} \label{subsec:res_meas}
This work aims to investigate the \jupfour{}/\jupthree{} properties of DYNAMO galaxies resolved on a $1-2$~kpc scale, and to relate this line ratio to the star formation rate surface density (\sfrsd{}) on the same scale. Thus, we describe here our method for extracting these measurements from the data. For each of our resolution and WCS matched ALMA and HST data sets (excluding SOFIA observations because they are unresolved), we define two sets of ``grids'' of circular, beam-sized apertures: one that is centered on the galaxy, and a second that is offset from the center by $0.5 \times$ the beam FWHM in both the $x$ and $y$ directions. This is to ensure that we cover the gaps of the first grid and results in measurements that are not entirely independent. Within each aperture, we measure the median brightness temperature of both the \jupthree{} and \jupfour{} lines from the integrated intensity maps and take their ratio. 

We measure the SFR surface density from our CO-matched H$\alpha$ observations. We perform aperture photometry within each ALMA beam-sized aperture in our two grids, described above, to obtain the H$\alpha$ flux (in electrons per second). We convert these fluxes to units of erg\,s$^{-1}$\,cm$^{-2}$\,\AA$^{-1}$, apply a correction for extinction by relating $A_{V}$ to $A_{\mathrm{H}\alpha}$ assuming the \citet{cardelli89} extinction law and the $A_{V}$ measurements from \citet{lenkic21}. \citet{bassett17} use Pa$\alpha$ observations from the OSIRIS instrument at Keck to make resolved extinction measurements in four DYNAMO galaxies. Their results show up to a magnitude difference in $A_{H\alpha}$; however, strong variation in the adaptive optics point spread function introduces significant systematic uncertainties in measuring the Pa$\alpha$ flux. Furthermore, \citet{bassett17} also find that the average $A_{H\alpha}$ in clump and non-clump regions are, within the uncertainties, consistent with one another (see their Table 3). This is consistent with the results of \citet{lenkic21}, who find that within a given DYNAMO galaxy, the extinction-sensitive color they measure shows little variation between clumps, and the clump colors are consistent with their host disks (see their Figures 5 and 8). For these reasons, we choose to adopt a single $A_{V}$ value for each galaxy. Finally, we calculate H$\alpha$ luminosities and convert them to SFRs using the \citet{hao11} calibration for a Kroupa IMF, constant star formation history, and age of 100~Myr (see their Table 2):

\begin{equation}
    \mathrm{SFR\;[M_{\odot}\,yr^{-1}]} = 5.53 \times 10^{-42} \times L_{\mathrm{H}\alpha} \; [\mathrm{erg\,s^{-1}}]
\end{equation}

\section{Results} \label{sec:results}
\subsection{\jupfour{}/\jupthree{} Line Ratios} \label{subsec:line_ratio}
\begin{figure*}
    \centering
    \includegraphics[height=4.3cm]{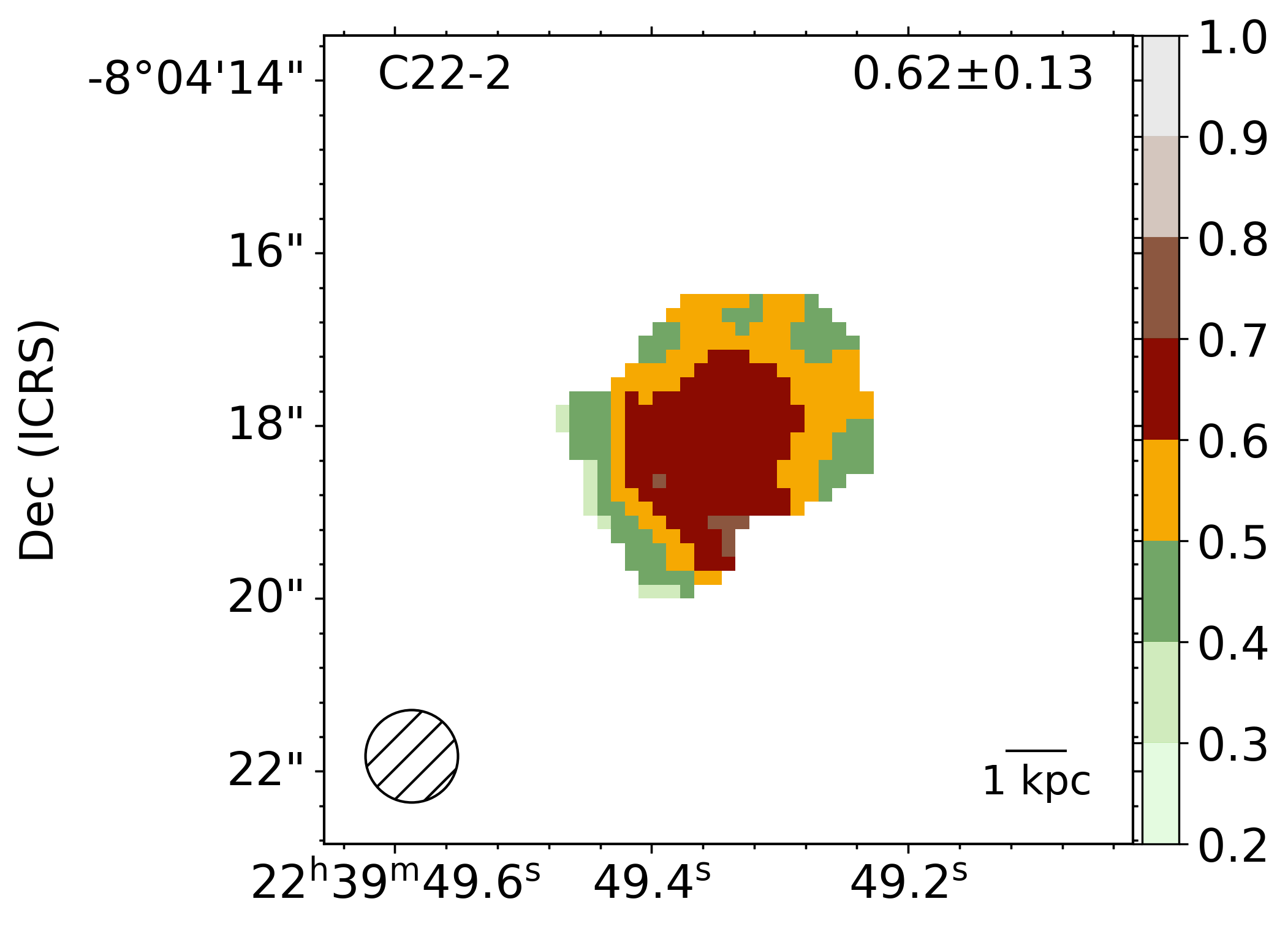}
    \includegraphics[height=4.3cm]{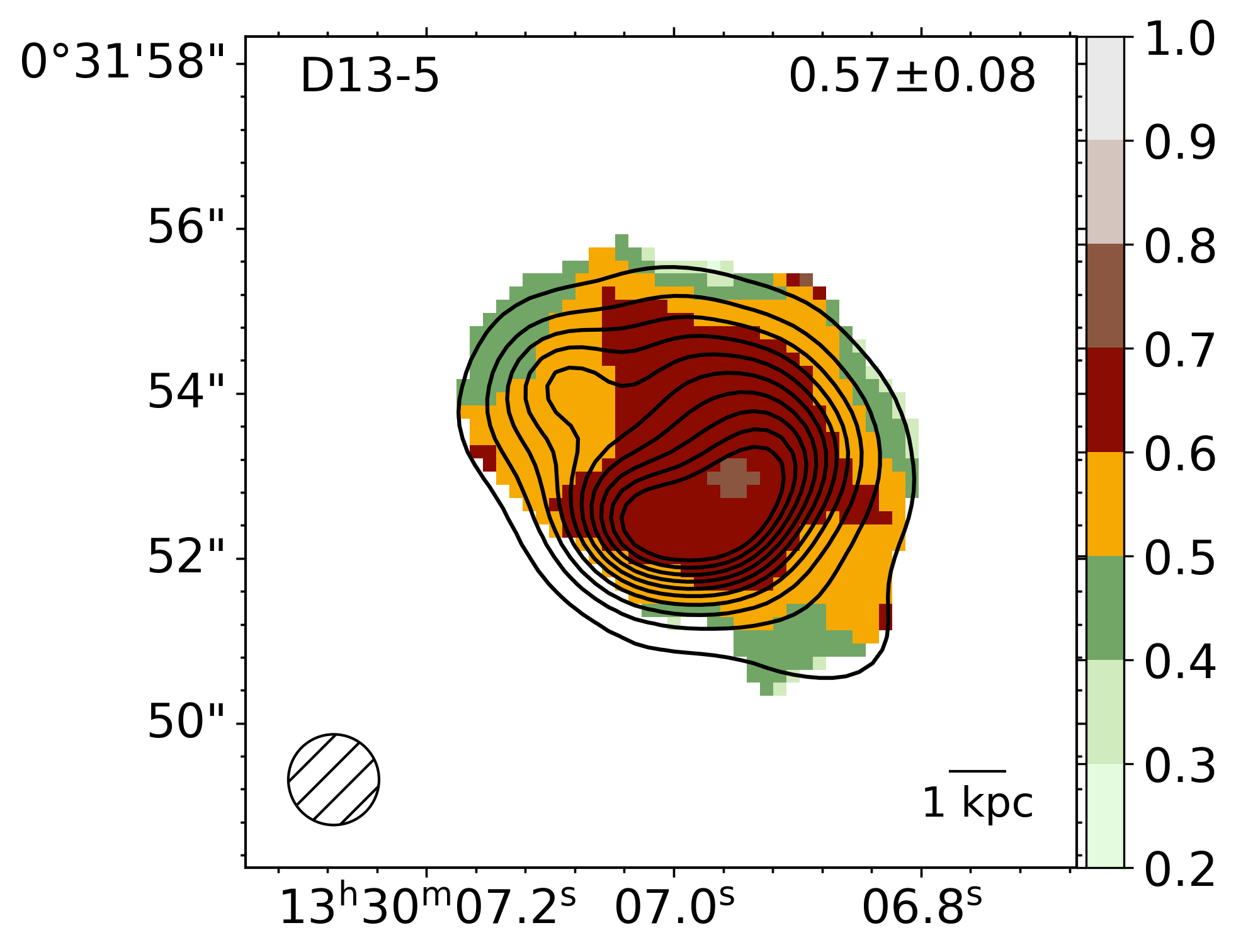}
    \includegraphics[height=4.3cm]{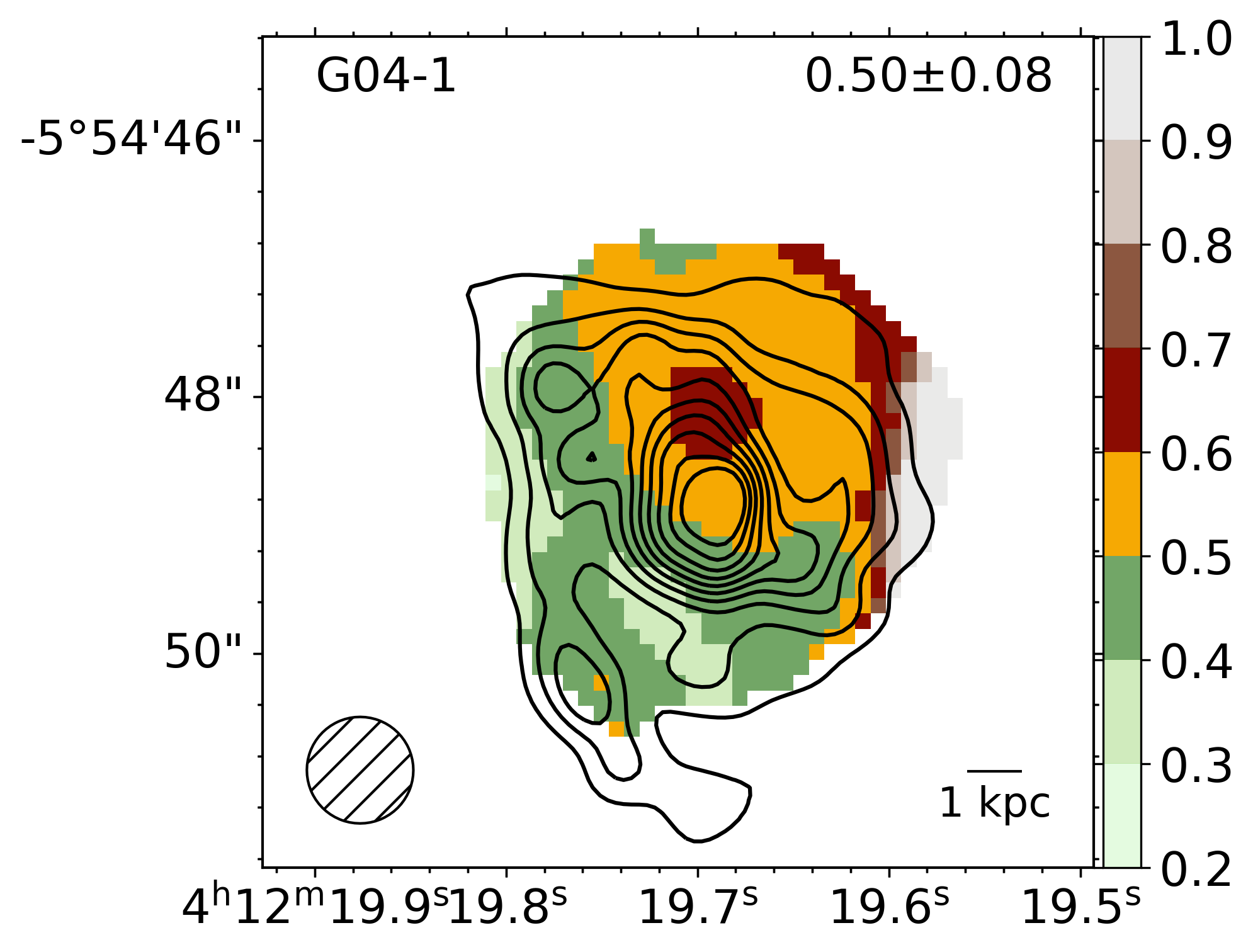}
    
    \includegraphics[height=4.6cm]{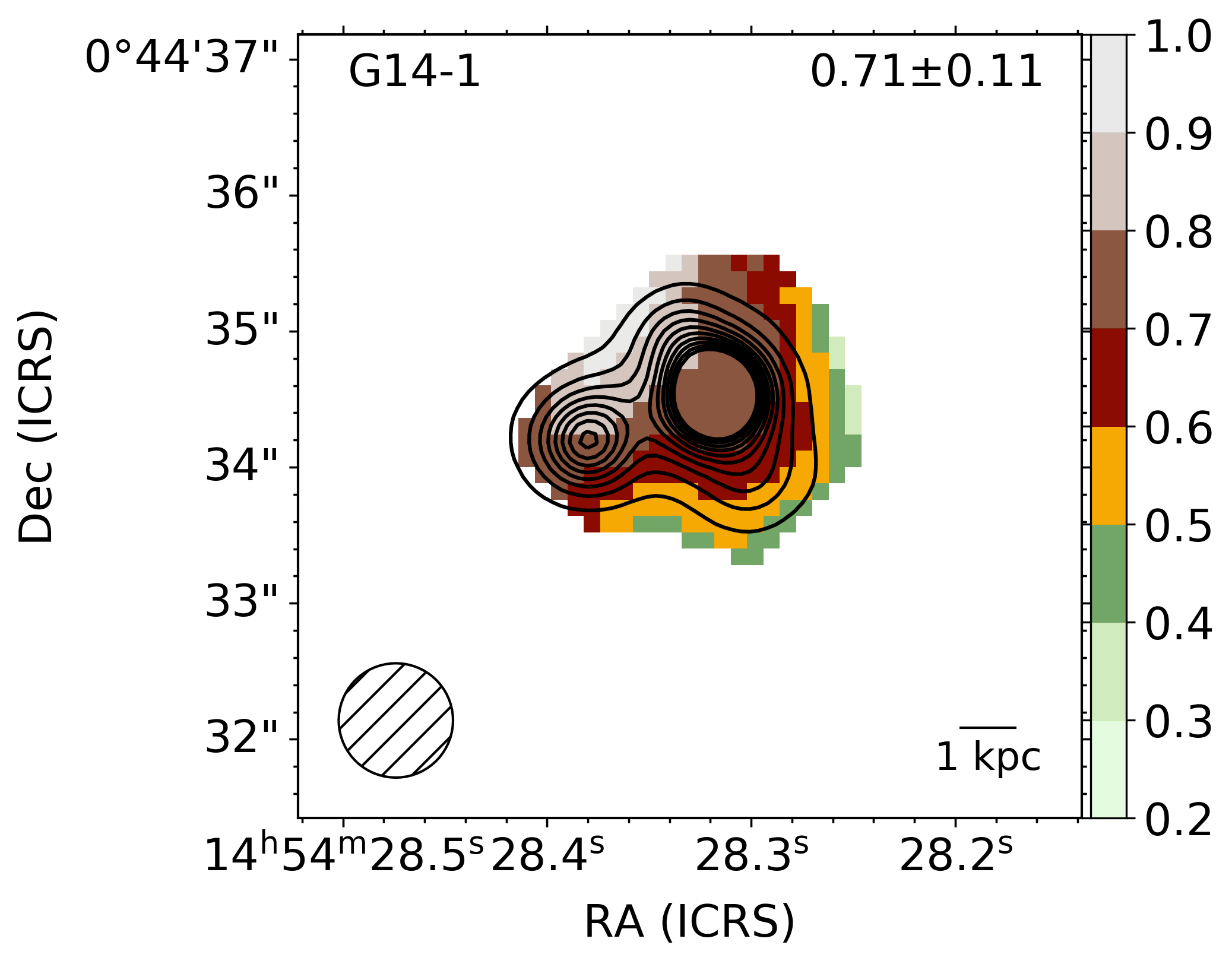}
    \includegraphics[height=4.6cm]{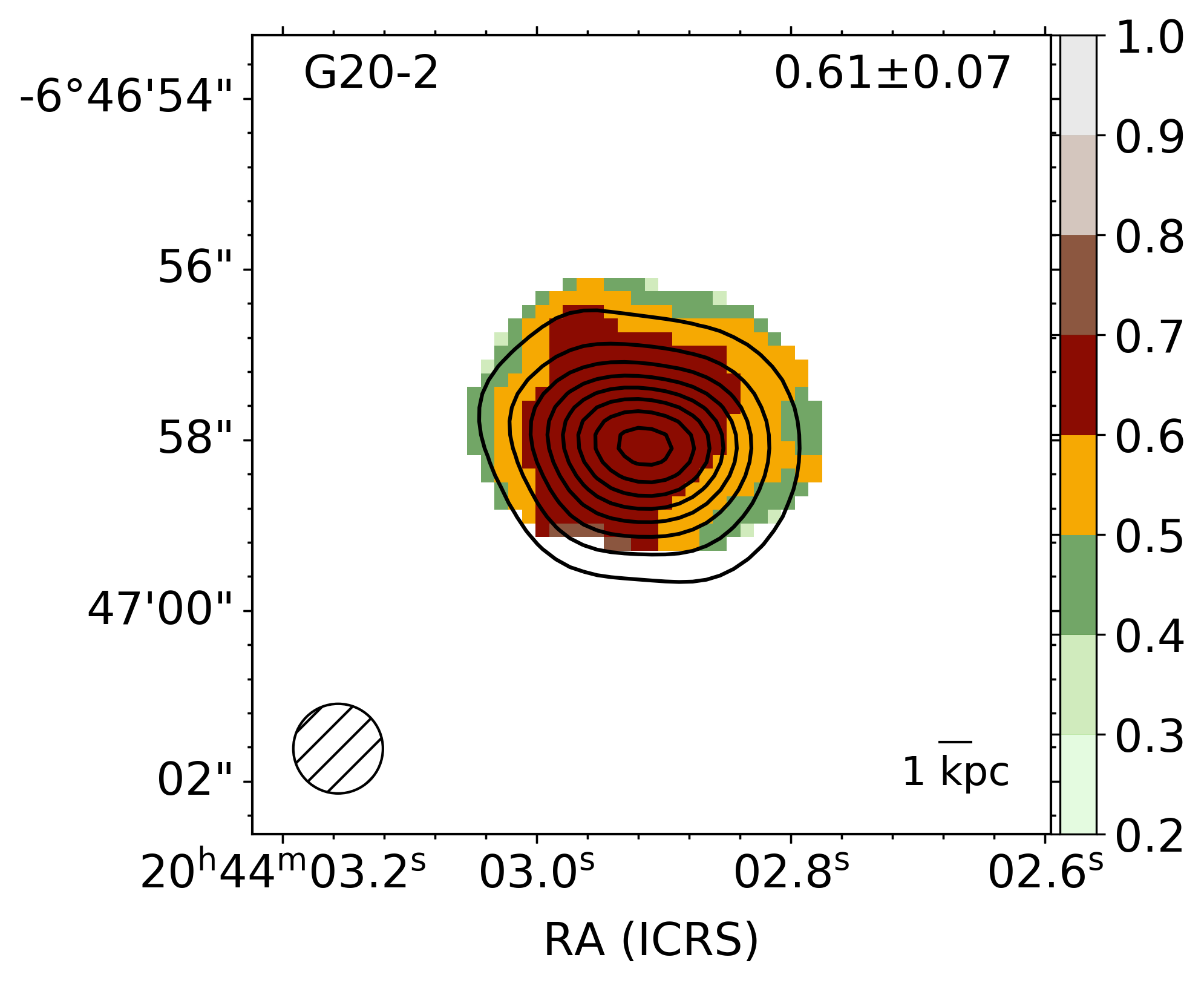}
    \includegraphics[height=4.6cm]{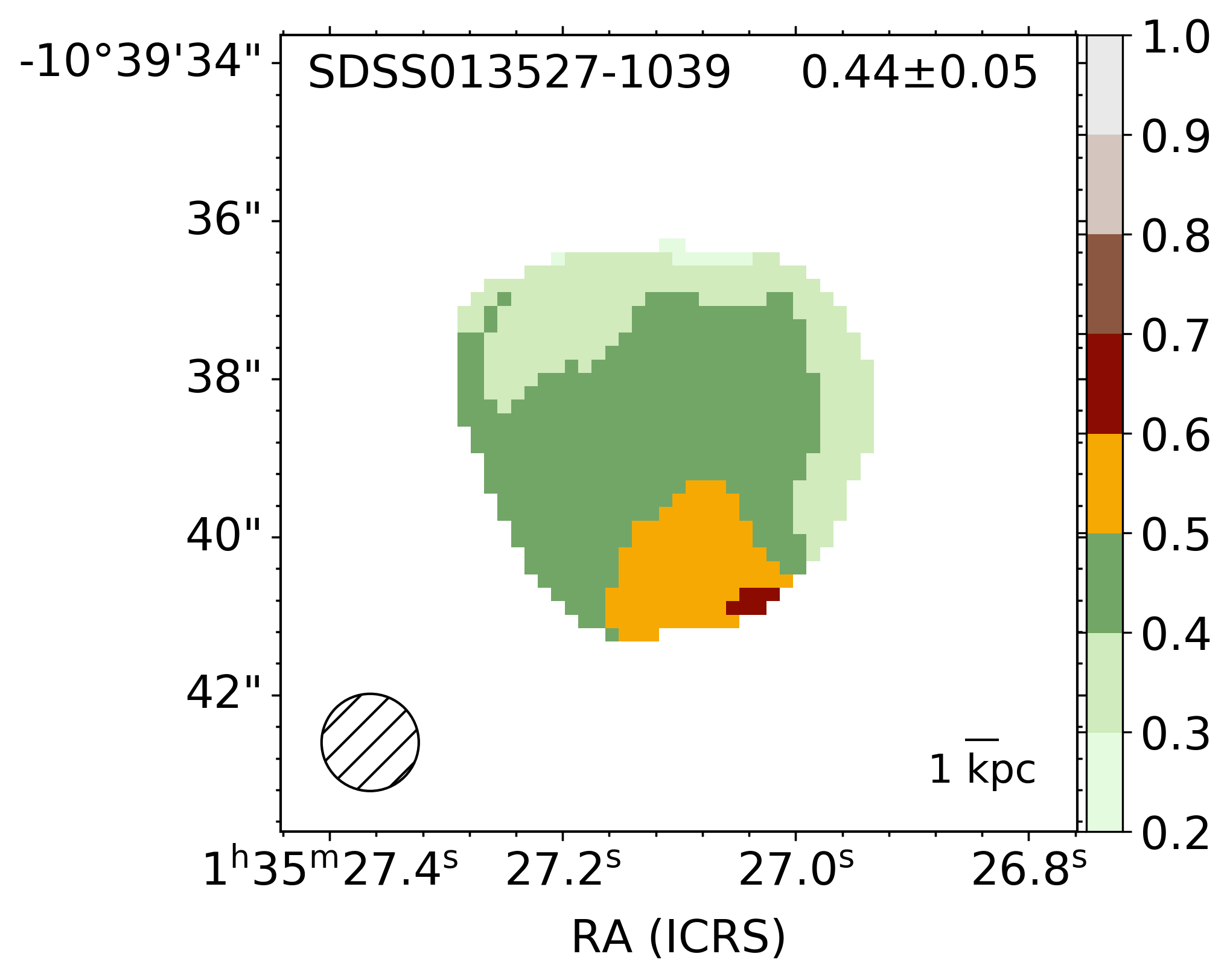}
    \caption{\jupfour{}/\jupthree{} line ratios (in brightness temperature units) measured from the pixel scale and resolution matched integrated intensity maps, integrated over the same velocity ranges. These maps show \jupfour{}/\jupthree{} only in regions where the line ratio S/N $\geq 3$. The black contours correspond to H$\alpha$ emission, where available, ranging from $1 -10 \sigma$ in increments of 1$\sigma$. The galaxy name is indicated in the top left corner of each panel, the median line ratios and their associated uncertainties are in the top right corners, and the black hatched circles in the bottom left corners indicate the circularized beam sizes. Finally, we show a 1~kpc scale bar in the bottom right corners. We see that galaxies generally have mildly varying line ratios within the regions where the uncertainties do not dominate, and that they lie typically around $0.4-0.7$, with the exception of DYNAMO G14-1 which has a stronger varying line ratio.}
    \label{fig:line_ratio}
\end{figure*}

In \S\ref{subsec:obs_alma}, we describe our process for matching our \jupfour{} and \jupthree{} observations and deriving integrated intensity maps. We adopt brightness temperature units, thus our integrated intensity maps have units of K\,km\,s$^{-1}$. To visually determine how this line ratio varies across each galaxy disk, if at all, we simply divide our \jupfour{} integrated intensity map by that of the \jupthree{}. This is what we present in Figure \ref{fig:line_ratio}, where the color scale indicates the ratio variations across each galaxy disk for which both line transitions were observed, and where the line ratio S/N $\geq 3$, and the black contours correspond to H$\alpha$ emission in the pixel scale and resolution matched HST observations. The contours span $1-10\sigma$ in increments of 1$\sigma$, where we take $\sigma$ to correspond to the rms of each HST observation calculated in galaxy emission-free regions. We note that there are no HST H$\alpha$ observations for DYNAMO C22-2 and SDSS 013527-1039. We derive uncertainties for the integrated intensity maps ($\sigma_{\mathrm{J \rightarrow J-1}}$) by summing in quadrature the rms of every channel over which we integrate, excluding line emission from the rms calculation, and multiplying by the channel width:

\begin{equation}
    \sigma_{\mathrm{J \rightarrow J-1}} = \Delta v\, \sqrt{\sum_{i}^{N} (\mathrm{rms}_{i})^{2}}
\end{equation}

\noindent where $\Delta v$ is the channel width, rms$_{i}$ is the rms of the $i^{th}$ channel, and N is the number of channels over which the emission is integrated. To obtain the final uncertainty on the line ratio per pixel, we propagate the integrated intensity uncertainties by taking:

\begin{equation}
    \sigma_{lr} = \frac{\mathrm{CO(4-3)}}{\mathrm{CO(3-2)}} \sqrt{\left(\frac{\sigma_{43}}{\mathrm{CO(4-3)}}\right)^{2} + \left(\frac{\sigma_{32}}{\mathrm{CO(3-2)}}\right)^{2}}
\end{equation}

\noindent which results in line ratio uncertainty maps.

From Figure \ref{fig:line_ratio}, we see that the line ratio for galaxies in our sample vary mildly across the disks, with typical values ranging from $R_{43} \sim 0.4 - 0.7$. However, galaxy DYNAMO G14-1 shows a strong gradient in the line ratio, with values approaching unity. The H$\alpha$ image of G14-1 in Figure \ref{fig:sum_obs} shows two bright clumps with a fainter ``stream'' connecting the two. The H$\alpha$ contours we over-plot in Figure \ref{fig:line_ratio} show that these two bright features with the connecting filament coincide with the elevated line ratio values and the strong gradient. This may be indicative of an interaction taking place; however, the H$\alpha$ kinematics of G14-1 show a rotating disk and no complex kinematics \citep{green14}. Overall, the line ratio maps we show in Figure \ref{fig:line_ratio} suggest a potential central enhancement in $R_{43}$. Such a central enhancement has been observed in the Milky Way and other nearby star-forming galaxies for \juptwo{}/\jupone{} \citep[see e.g.,][]{sakamoto97,sawada01,leroy09,leroy13,denBrok21}. To verify this, we separate pixels that are located within the central kpc of each galaxy from pixels that lie outside this region, and compare the median line ratios. Indeed, we find enhanced \jupfour{}/\jupthree{} values in the central kpc of all galaxies in Figure \ref{fig:line_ratio}, except for G04-1 and G14-1, on the order of $\sim 10$\% (see Table \ref{tab:lrs}). Finally, Figure \ref{fig:line_ratio} shows in particular for galaxy G04-1, variations in $R_{43}$ between the spiral arm and inter-arm region, a trend also observed for \juptwo{}/\jupone{} in M51 \citep{koda12}.

\begin{figure}
    \centering
    \includegraphics[width=\columnwidth]{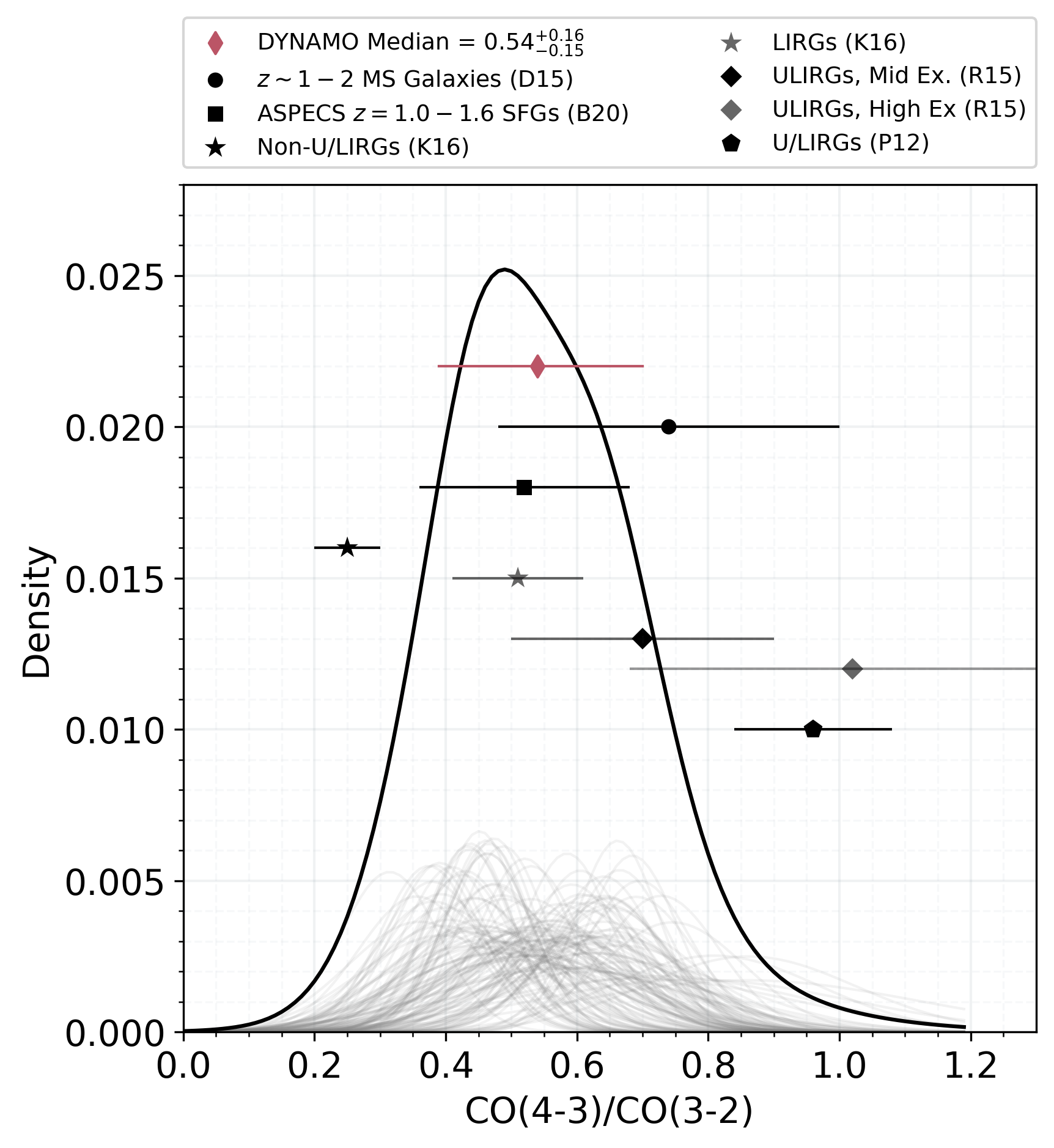}
    \caption{Global PDF for the resolved \jupfour{}/\jupthree{} line ratio measurements. We construct the PDF by modeling each line-of sight $R_{43}$ measurement (where S/N $\geq 3$) as a Gaussian whose width is the line ratio uncertainty. We show these individual Gaussians as light grey lines (not to scale); summing them and normalizing the area of the resulting Gaussian to unity results in the solid black line shown here. From the cumulative distribution function, we infer a median line ratio of $R_{43} = 0.54$. For comparison, we include estimates from the literature: $R_{43} = 0.74 \pm 0.26$ for three $z \sim 1.5$ main sequence star forming galaxies \citep[black circle;][]{daddi15}, $R_{43} = 0.52 \pm 0.16$ in eight star-forming galaxies at $z = 1.0-1.6$ \citep[black square;][]{boogaard20}, $R_{43} = 0.25 \pm 0.05, 0.51 \pm 0.10$ for non-U/LIRGs with $L_{\mathrm{FIR}} = 10^{10}\,\mathrm{L}_{\odot}$ and U/LIRGs with $L_{\mathrm{FIR}} = 10^{11}\,\mathrm{L}_{\odot}$ respectively \citep[black stars;][]{kamenetzky16}, $R_{43} = 0.70, 1.02$ for mid- and high-excitation ULIRGs respectively \citep[black diamonds;][]{rosenberg15}, and $R_{43} = 0.96 \pm 0.12$ for LIRGs \citep[black pentagon;][]{papadopoulos12}.}
    \label{fig:lr_kde}
\end{figure}

Next, we perform $\sim 1-2$~kpc sized sightline measurements of the line ratio across the disk of each galaxy, as described in \S\ref{subsec:res_meas}, to characterize the typical line ratio we measure across the sample and the magnitude of the spread. To this end, we construct a global probability density function (PDF) by modeling each beam-averaged line ratio measurement with a kernel density estimate (KDE). We construct the individual KDEs by modeling each beam-averaged line ratio measurement as a one-dimensional Gaussian with centroid corresponding to the measured line ratio and with width equal to the line ratio uncertainty. The area of each Gaussian is normalized to unity, then we sum all Gaussians to produce a final global PDF \citep[see for example \S 4 and Figure 5 of][]{levy18}. This is what we show in  Figure \ref{fig:lr_kde}. From this, we find that the median line ratio and 68\% confidence interval for DYNAMO galaxies are: $R_{43} = 0.54${\raisebox{0.5ex}{\tiny$^{+0.16}_{-0.15}$}}. These values are taken at the 15.9, 50, and 84.1 percentiles of the cumulative distribution function of the PDF.

For comparison, we compile estimates of the \jupfour{} to \jupthree{} line ratio from the literature and include these in Figure \ref{fig:lr_kde}. We describe our derivation of all line ratios we compile from the literature in Appendix \ref{app:lrs}, and we summarize them along with the median line ratios we measure for each DYNAMO galaxy individually, and the median line ratio for the entire DYNAMO sample studied here in Table \ref{tab:lrs}.

\begin{deluxetable}{lcc}
\tablecaption{CO(4$-$3)/CO(3$-$2) Line Brightness Temperature Ratios Compiled from the Literature Compared to DYNAMO} 
\label{tab:lrs}
\tablewidth{700pt}
\tabletypesize{\footnotesize}
\tablehead{
    \colhead{Object(s)}        &
	\colhead{Line Ratio}       &
	\colhead{Reference}        
} 
\startdata
C22-2 & 0.62 $\pm$ 0.13 & This work \\
C22-2 ($\leq 1$~kpc) & 0.60 $\pm$ 0.10 & This work \\
C22-2 ($> 1$~kpc) & 0.52 $\pm$ 0.08 & This work \\
D13-5 & 0.57 $\pm$ 0.08 & This work \\
D13-5 ($\leq 1$~kpc) & 0.60 $\pm$ 0.08 & This work \\
D13-5 ($> 1$~kpc) & 0.56 $\pm$ 0.09 & This work \\
G04-1 & 0.50 $\pm$ 0.08 & This work \\
G04-1 ($\leq 1$~kpc) & 0.48 $\pm$ 0.20 & This work \\
G04-1 ($. 1$~kpc) & 0.52 $\pm$ 0.16 & This work \\
G14-1 & 0.71 $\pm$ 0.11 & This work \\
G14-1 ($\leq 1$~kpc) & 0.70 $\pm$ 0.13 & This work \\
G14-1 ($> 1$~kpc) & 0.71 $\pm$ 0.16 & This work \\
G20-2 & 0.61 $\pm$ 0.07 & This work \\
G20-2 ($\leq 1$~kpc) & 0.62 $\pm$ 0.10 & This work \\
G20-2 ($> 1$~kpc) & 0.57 $\pm$ 0.10 & This work \\
SDSS J013527.10-103938.6 & 0.44 $\pm$ 0.05 & This work \\
SDSS J013527.10-103938.6 ($\leq 1$~kpc) & 0.46 $\pm$ 0.06 & This work \\
SDSS J013527.10-103938.6 ($> 1$~kpc) & 0.42 $\pm$ 0.06 & This work \\
DYNAMO all & $0.54${\raisebox{0.5ex}{\tiny$^{+0.16}_{-0.15}$}} & This work \\
\hline
$z = 1.5$ MS Galaxies & 0.74 $\pm$ 0.26 & D15 \\
ASPECS $z = 1.0-1.6$ SFGs & 0.52 $\pm$ 0.16 & B20 \\
G1700-MD94 one component & $0.92 \pm 0.18$ & HB22 \\
G1700-MD94 two component & $0.77 \pm 0.15$ & HB22 \\
non-U/LIRGs ($L_{\mathrm{FIR}} = 10^{10} L_{\odot}$) & $0.25 \pm 0.05$ & K16 \\
LIRGs ($L_{\mathrm{FIR}} = 10^{11} L_{\odot}$) & $0.51 \pm 0.10$ & K16 \\
LIRGs & 1.23 $\pm$ 0.38 & P12 \\
ULIRGs low CO excitation & 1.08 & R15 \\
ULIRGs mid CO excitation & 0.70 & R15 \\
ULIRGs high CO excitation & 1.02 & R15 \\
\enddata
\end{deluxetable}

This comparison reveals that the \jupfour/\jupthree\ line ratio of non-ULIRGs from \citet{kamenetzky16} (local galaxies with $L_{\mathrm{FIR}} \leq 6 \times 10^{10}\,L_{\odot}$) is much lower and incompatible with what we find in our DYNAMO sample. In contrast, the U/LIRG line ratio estimate from \citet{kamenetzky16} for $L_{\mathrm{FIR}} = 10^{11}\,L_{\odot}$ is in much better agreement with what we find across the DYNAMO sample. Likewise, the \jupfour/\jupthree\ line ratios measured in main-sequence galaxies at $z \sim 1-2$ \citep{daddi15,boogaard20,henriquez-brocal22} are, within the uncertainties, consistent with DYNAMO. In particular, the eight star-forming galaxies at $z = 1.0-1.6$ from the ALMA Spectroscopic Survey \citep[ASPECS;][]{boogaard20}, are an especially good match to the $R_{43}$ we measure across our sample. DYNAMO galaxies lie on the star formation main-sequence at $z \sim 2$ \citep{fisher19} and have gas fractions and velocity dispersions that are more similar to main-sequence galaxies of that epoch than local ones. Therefore, this result is consistent with lines of evidence that indicate DYNAMO galaxies are local analogues of high$-z$ main-sequence systems. ULIRG samples \citep{rosenberg15} have much larger line ratios than we observe in DYNAMO, and this too is consistent with previous observations. Using \textit{Herschel} PACS+SPIRE observations of five DYNAMO galaxies, \citet{white17} found that despite their large FIR luminosities, $L_{\mathrm{FIR}} > 10^{11}\,\mathrm{L}_{\odot}$, these galaxies have much lower dust temperatures ($\sim 30$~K) than ULIRGs. Therefore, unlike ULIRGs, the star formation in DYNAMO galaxies is more distributed throughout the disks; thus, colder dust temperatures would be expected and likewise lower \jupfour/\jupthree\ line ratios.

\subsection{Relating High$-$J CO to CO(1$-$0)} \label{subsec:co10}
\begin{deluxetable*}{lccccccc}
\tablecaption{Galaxy Integrated CO(3$-$2)/CO(1$-$0) and CO(4$-$3)/CO(1$-$0) Line Ratios and Model Predictions} 
\label{tab:lrs_co10}
\tablewidth{700pt}
\tabletypesize{\normalsize}
\tablehead{
    \colhead{Galaxy}                                         &
	\colhead{$L'_{CO(1-0)}$}                                 &
	\colhead{$L'_{CO(3-2)}$}                                 &
	\colhead{$L'_{CO(4-3)}$}                                 &
	\colhead{$R_{31}$}                                       &
	\colhead{$R_{41}$}                                       &
	\colhead{$R_{31}$}                                       &
	\colhead{$R_{41}$}                                       \\
	\colhead{}                                               &
	\multicolumn{3}{c}{(10$^{9}$~K\,km\,s$^{-1}$\,pc$^{2}$)} &
	\multicolumn{2}{c}{Observed}                             &
	\multicolumn{2}{c}{Predicted}                            
} 
\startdata
C13-1 & $1.91 \pm 0.05$ & $0.47 \pm 0.05$ & $\cdots$ & $0.40 \pm 0.05$ & $\cdots$ & $\cdots$ & $\cdots$ \\
C22-2 & $0.66 \pm 0.05$ & $0.47 \pm 0.04$ & $0.23 \pm 0.02$ & $0.71 \pm 0.08$ & $0.35 \pm 0.04$ & $\cdots$ & $\cdots$ \\
D13-5 & $2.69 \pm 0.08$ & $1.48 \pm 0.03$ & $0.87 \pm 0.03$ & $0.55 \pm 0.02$ & $0.32 \pm 0.01$ & 0.63 & 0.39 \\
D15-3 & $3.02 \pm 0.06$ & $\cdots$ & $0.49 \pm 0.02$ & $\cdots$ & $0.16 \pm 0.01$ & 0.57 & 0.32 \\
G04-1 & $5.41 \pm 0.39$ & $2.90 \pm 0.10$ & $1.54 \pm 0.08$ & $0.54 \pm 0.04$ & $0.28 \pm 0.02$ & 0.56 & 0.31 \\
G08-5 & $2.29 \pm 0.26$ & $1.83 \pm 0.08$ & $\cdots$ & $0.80 \pm 0.10$ & $\cdots$ & 0.57 & 0.32 \\
G14-1 & $1.59 \pm 0.19$ & $0.77 \pm 0.08$ & $0.427 \pm 0.001$ & $0.48 \pm 0.08$ & $0.27 \pm 0.03$ & 0.56 & 0.31 \\
G20-2 & $1.68 \pm 0.19$ & $0.97 \pm 0.03$ & $0.56 \pm 0.03$ & $0.58 \pm 0.07$ & $0.33 \pm 0.04$ & 0.61 & 0.36 \\
SDSS 013527-1039 & $3.45 \pm 0.16$ & $1.48 \pm 0.05$ & $0.65 \pm 0.02$ & $0.43 \pm 0.02$ & $0.19 \pm 0.01$ & $\cdots$ & $\cdots$ \\
\enddata
\end{deluxetable*}

We make use of existing \jupone{} measurements from the PdBI and NOEMA \citep[angular resolution $\sim 5-10$\arcsec{};][]{fisher14,white17,fisher19} to derive the \jupfour{}/\jupone{} and \jupthree{}/\jupone{} line ratios across our sample. We measure the total \jupfour{} and \jupthree{} fluxes by summing all pixels with S/N $\geq 3$ in our integrated intensity moment maps, then scaling by the number of pixels per beam. We then convert the total fluxes to luminosities ($L'$; K\,km\,s$^{-1}$\,pc$^{2}$) using equation 3 in \citet{solomon97}. We present these results in Table \ref{tab:lrs_co10}. 

\begin{figure}
    \centering
    \includegraphics[width=\columnwidth]{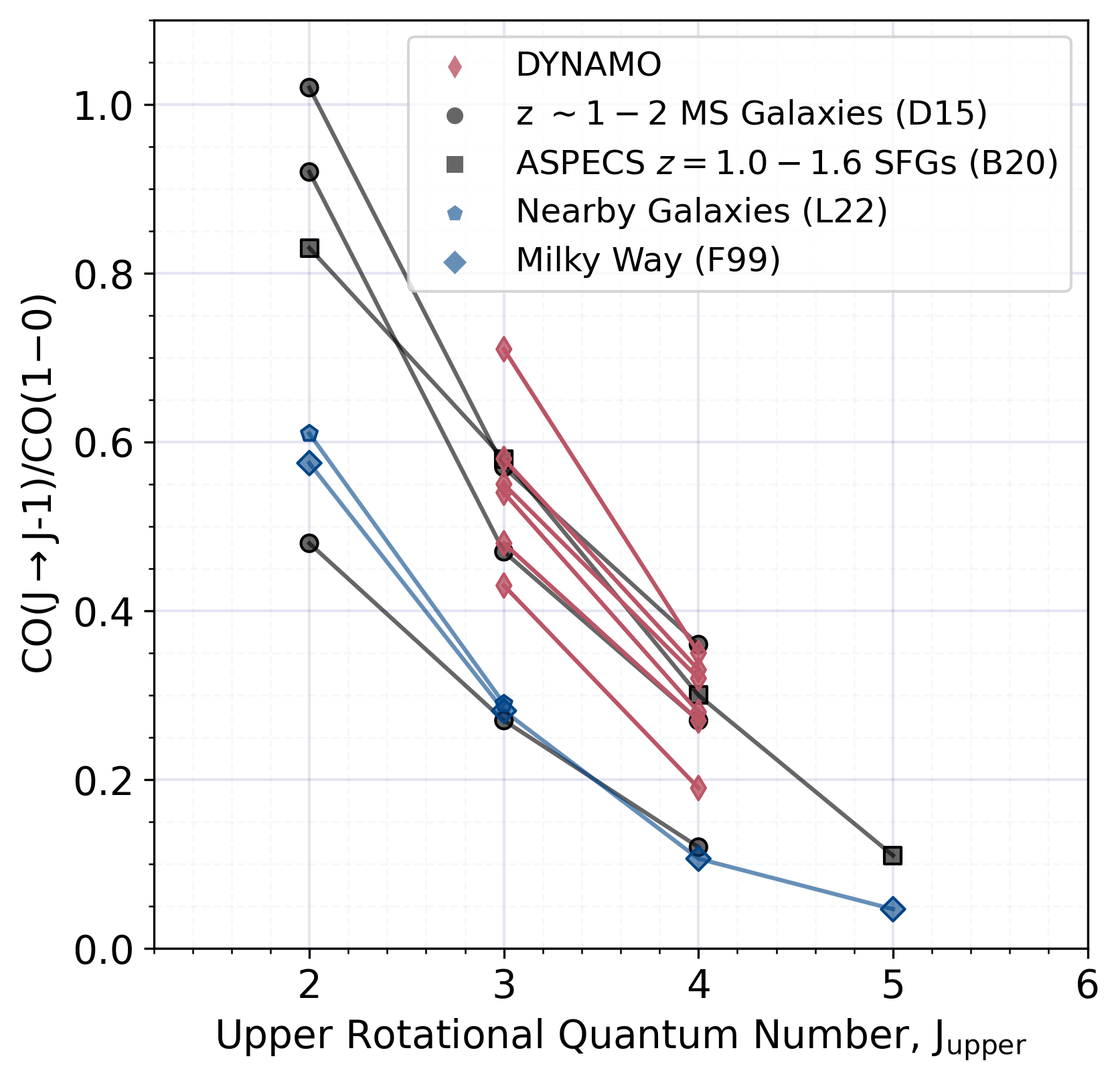}
    \caption{CO ladders normalized to \jupone{} in integrated brightness temperature units, for DYNAMO galaxies (red small diamond), $z \sim 1-2$ main-sequence BzK galaxies \citep[black circles;][]{daddi15}, $z = 1.0-1.6$ star-forming galaxies from ASPECS \citep[black squares][]{boogaard20} nearby star forming galaxies \citep[blue pentagons;][]{leroy22}, and the Milky Way inner disk \citep[blue large diamonds;][]{fixsen99}. DYNAMO line ratios are consistent with $z \sim 1-2$ star-forming galaxies, while the nearby star-forming galaxies and the Milky Way show overall lower CO excitation. One galaxy from the sample of \citet{daddi15} (BzK-16000) is more consistent with nearby galaxies and the Milky Way than with DYNAMO. BzK-16000 is more evolved and has no massive clumps.}
    \label{fig:sled}
\end{figure}

We find line ratio values across our sample that range from $R_{31} \sim 0.4-0.8$, with a mean (median) $R_{31} = 0.56\, (0.55)$ and $R_{41} \sim 0.2 - 0.4$, with a mean (median) $R_{41} = 0.27\, (0.28)$. Our $R_{31}$ result is consistent with multiple studies of CO excitation in $z \sim 1-3$ star-forming galaxies: \citet{daddi15} found that the brightness temperature line ratio of \jupthree{} to \jupone{} ranges from $R_{31} \sim 0.4 - 0.6$, with an average $R_{31} = 0.42 \pm 0.07$, for their three star-forming BzK $z \sim 1.5$ galaxies, \citet{dessauges-zavadsky15} find $R_{31} = 0.57 \pm 0.15$ for five lensed star-forming galaxies (SFR $< 40$~M$_{\odot}$\,yr$^{-1}$) at $z \sim 1.5$ galaxies, \citet{riechers20} find $R_{31} = 0.84 \pm 0.26$ for six galaxies at $z \sim 2-3$, \citet{birkin21} find $R_{31} = 0.63 \pm 0.12$ for a large sample of SMGs at $z \sim 1.2-4.8$, and \citet{harrington21} find $R_{31} = 0.69 \pm 0.12$ for 24 dusty star forming galaxies at $1 < z < 3$. However, we note that \citet{bolatto15} find $R_{31} \sim 1$ for two main-sequence galaxies at $z \sim 2$; however, one of the two galaxies is classified as an AGN, and the other may host an weak AGN. In contrast, \citet{leroy22} analyze the global $R_{31}$ for nearby normal star-forming galaxies and find a mean (median) of $R_{31} = 0.30\, (0.29)$, which are lower and inconsistent with the DYNAMO results, but similar to the Milky Way \citep{fixsen99}. We illustrate this comparison in Figure \ref{fig:sled} where we plot the brightness temperature ratios of DYNAMO galaxies as a function of upper-J number, along with the ratios of $z \sim 1-2$ star-forming galaxies \citep{daddi15,boogaard20}, nearby star-forming galaxies \citep{leroy22}, and the Milky Way inner disk \citep{fixsen99}. We can see that the ratios of nearby galaxies and the Milky Way are incompatible with those of DYNAMO. The CO SLED of the ASPECS galaxies and two of the three BzK galaxies are in agreement with DYNAMO, while the third galaxy \citep[referred to as BzK-16000 in][]{daddi15} shows overall lower line ratios. Interestingly, \citet{daddi15} describe this galaxy as the most evolved in their sample, with no massive clumps.

\subsection{DYNAMO SEDs and \CII{} Emission} \label{subsec:sed}
We extract background subtracted 89, 155, and 216~\micron{} fluxes for four DYNAMO galaxies, including D15-3 which overlaps with our ALMA sample, from our HAWC+ SOFIA observations using the \textsc{Photutils} software \citep{larry_bradley_2021_5525286}. We define the flux extraction apertures to correspond to the FWHM beam size of each corresponding HAWC+ band, while we define the background annuli to have an inner radius equal to $5 \times$\,beam FWHM and an outer radius of $7 \times$\,beam FWHM (see Figure \ref{fig:hawc_obs} in Appendix \ref{app:sofia}). We record these flux measurements in Table \ref{tab:sofia}. To fit the SED, we combine the HAWC+ fluxes with WISE measurements at 22~\micron{} \citep[which is not contaminated by line emission and traces the warm dust continuum;][]{cluver17}, and use the modified blackbody (MBB) SED fitting tool \textsc{mbb\_emcee}\footnote{\href{https://github.com/aconley/mbb\_emcee}{https://github.com/aconley/mbb\_emcee}}, described in \citet{riechers13} and \citet{dowell14}. The MBB is joined to a power law of the form $\nu^{\alpha}$ at short wavelengths. \textsc{mbb\_emcee} fits the dust temperature, T$_{\mathrm{d}}$, the extinction curve power law slope, $\beta$, the power law slope of the blue side, $\alpha$, the wavelength where the optical depth reaches one, $\lambda_{0}$, and the normalization. We impose a prior on $\beta$ to constrain it between $1.5-2$, and leave all other parameters unconstrained. We record the resulting fit parameters and total infrared luminosity ($8-1000$~\micron{}; TIR) in Table \ref{tab:sofia}.

\begin{deluxetable*}{lccccccc}
\tablecaption{SOFIA [CII] and IR Measurements} 
\label{tab:sofia}
\tablewidth{700pt}
\tabletypesize{\normalsize}
\tablehead{
    \colhead{Galaxy}                            &
    \colhead{22.2~\micron{}}                    &
	\colhead{89~\micron{}}                      &
	\colhead{155~\micron{}}                     &
	\colhead{216~\micron{}}                     &
	\colhead{T$\mathrm{_{d}}$}                  &
	\colhead{TIR}                               &
	\colhead{log$_{10}$ L$_{\mathrm{[CII]}}$}  \\
	\colhead{}                                  &
	\colhead{(mJy)}                             &
	\colhead{(mJy)}                             &
	\colhead{(mJy)}                             &
	\colhead{(mJy)}                             &
	\colhead{(K)}                               &
	\colhead{(10$^{10}$~L$_{\odot}$)}             &
	\colhead{(erg\,s$^{-1}$)}
} 
\startdata
B08-3 & $\cdots$ & $\cdots$ & $\cdots$ & $\cdots$ & $\cdots$ & $\cdots$ & 41.82 $\pm$ 0.46 \\
D10-4 & $\cdots$ & $\cdots$ & $\cdots$ & $\cdots$ & $\cdots$ & $\cdots$ & 41.91 $\pm$ 0.46 \\
D14-1 & 12.3 $\pm$ 3.5 & 148 $\pm$ 19 & 423 $\pm$ 47 & 209 $\pm$ 25 & 25.94{\raisebox{0.5ex}{\tiny$^{+5.10}_{-4.88}$}} & 3.48{\raisebox{0.5ex}{\tiny$^{+0.69}_{-0.78}$}} & $\cdots$ \\
D15-3 & 8.1 $\pm$ 2.8 & 282 $\pm$ 33 & 393 $\pm$ 44 & 685 $\pm$ 73 & 29.56{\raisebox{0.5ex}{\tiny$^{+3.06}_{-2.89}$}} & 4.26{\raisebox{0.5ex}{\tiny$^{+0.52}_{-0.49}$}} & 41.74 $\pm$ 0.46 \\
F08-2 & $\cdots$ & 522 $\pm$ 57 & 326 $\pm$ 37 & 563 $\pm$ 61 & 46.15{\raisebox{0.5ex}{\tiny$^{+7.66}_{-6.72}$}} & 6.27{\raisebox{0.5ex}{\tiny$^{+1.30}_{-1.34}$}} & 41.95 $\pm$ 0.46 \\
F09-1 & $\cdots$ & $\cdots$ & $\cdots$ & $\cdots$ & $\cdots$ & $\cdots$ & 42.20 $\pm$ 0.46 \\
F12-4 & 15.3 $\pm$ 3.9 & 224 $\pm$ 27 & 279 $\pm$ 32 & 594 $\pm$ 64 & 23.43{\raisebox{0.5ex}{\tiny$^{+7.65}_{-7.05}$}} & 4.11{\raisebox{0.5ex}{\tiny$^{+0.74}_{-1.02}$}} & 42.21 $\pm$ 0.46 \\
\enddata
\end{deluxetable*}

\begin{figure*}
    \centering
    \includegraphics[width=\columnwidth]{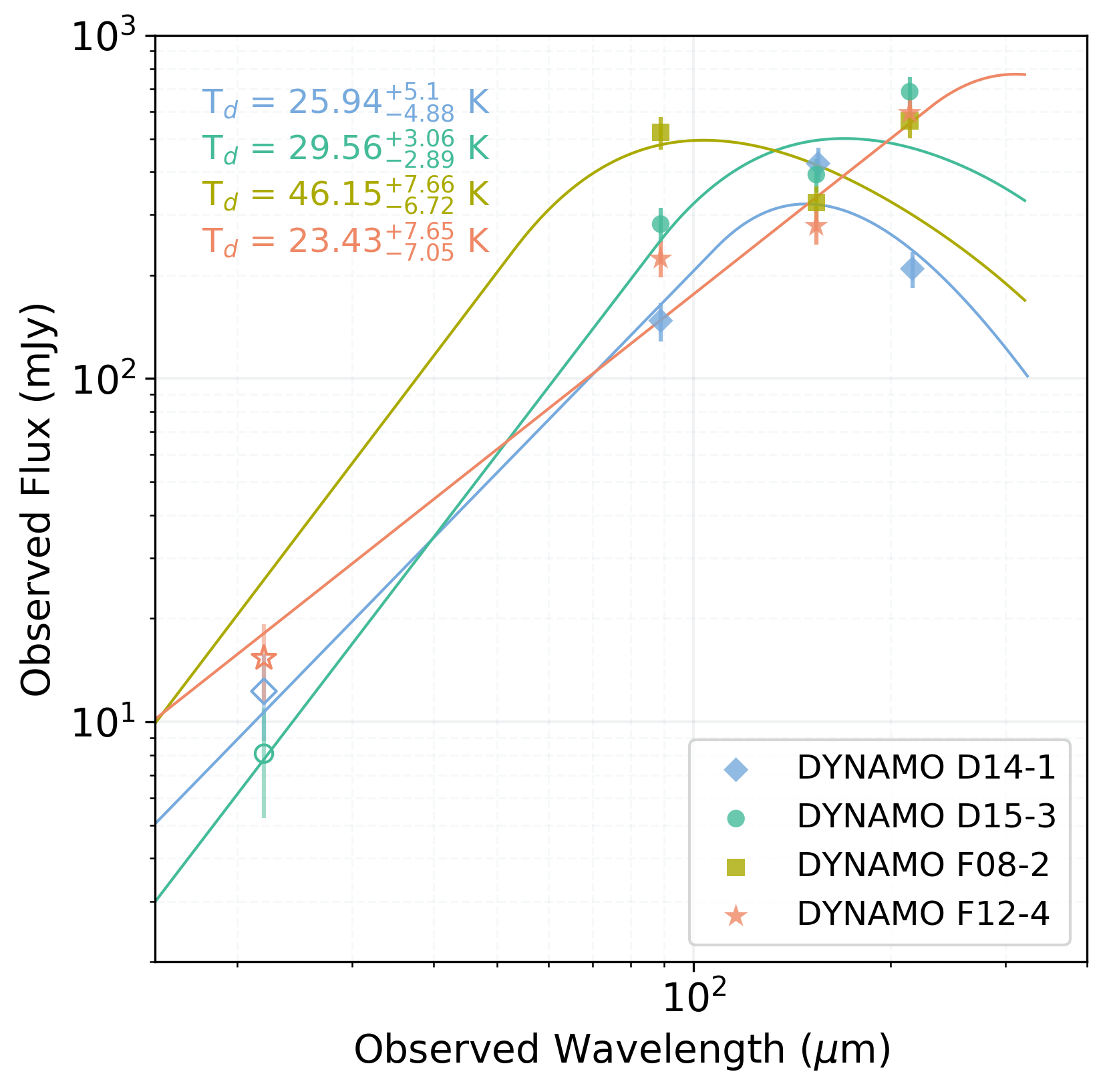}
    \includegraphics[width=\columnwidth]{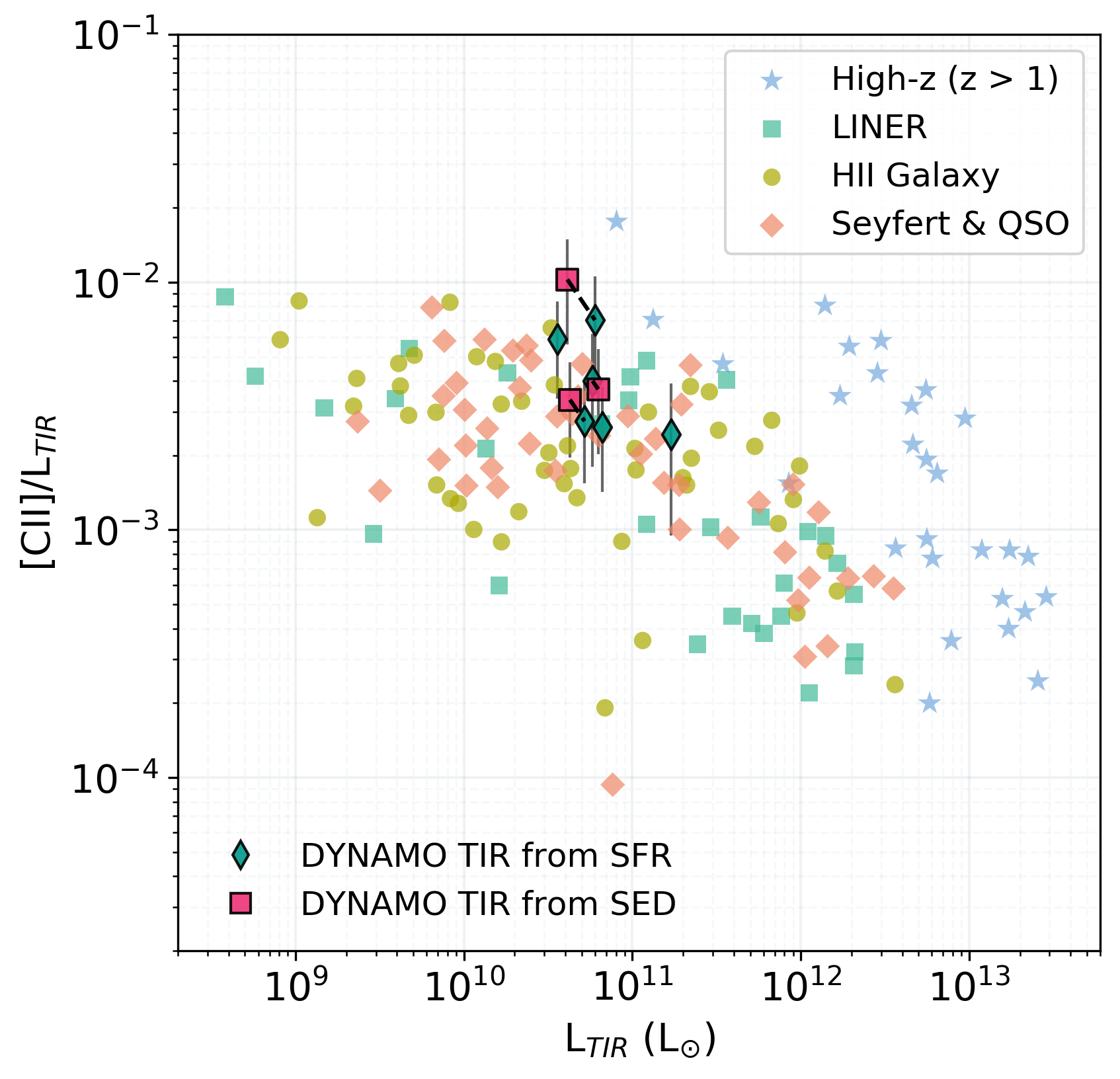}
    \caption{\textit{Left:} Spectral energy distribution of galaxies DYNAMO D14-1, D15-3, F08-2, and F12-4 based on fluxes from SOFIA HAWC+ (colored symbols) and WISE observations (open colored symbols). The solid colored lines are the resulting SED fits using \textsc{mbb\_emcee}, with the corresponding dust temperatures appearing in the top left corner. For DYNAMO D14-1 and D15-3, the dust temperatures derived from the HAWC+ measurements are consistent with those derived by \citet{white17} using \textit{Herschel} PACS and SPIRE photometry. \textit{Right:} The \CII{}-to-TIR ratio as a function of TIR. The colored symbols with black outlines correspond to DYNAMO measurements; the magenta symbols have TIR measurements derived from SED fitting, while the teal symbols have TIR estimated from SFRs. For galaxies where we have both SED measurements and SFRs, we link the data points via a black dashed line. Grey error bars are the assumed 40\% calibration uncertainty for the \CII{} observations and TIR uncertainties propagated through the ratio. We compare our DYNAMO measurements to those of \citet{herrera-camus18a} and find that DYNAMO galaxies do not show a deficit of \CII{} emission, consistent with their cooler dust temperatures.}
    \label{fig:all_sed}
\end{figure*}

We present the resulting SEDs in the left panel of Figure \ref{fig:all_sed}, where the filled colored data points represent fluxes from the HAWC+ (three longest wavelength data points), open colored data points represent WISE bands (shortest wavelength point) for each of the four galaxies, and the matching colored line represents the SED fit for that galaxy. The dust temperatures derived from these SEDs are shown in the upper left corner. For DYNAMO D14-1 and D15-3, the resulting dust temperatures are $T_{d} = 25.94${\raisebox{0.5ex}{\tiny$^{+5.10}_{-4.88}$}}~K and $T_{d} = 29.56${\raisebox{0.5ex}{\tiny$^{+3.06}_{-2.89}$}}~K respectively, consistent with the dust temperature measurements of $T_{d} = 28.09 \pm 0.86$~K and $T_{d} = 25.64 \pm 0.52$~K from SED fitting of \textit{Herschel} PACS and SPIRE observations by \citet{white17}.

The SEDs provide us with estimates of TIR for four out of the seven galaxies in the SOFIA sample. We combine these measurements with the SOFIA FIFI-LS observations to explore the “\CII{}-deficit”: the observed decreasing fraction of \CII{} emission with respect to TIR in increasingly more infrared luminous objects \citep[see e.g.,][]{malhotra01,brauher08,smith17,herrera-camus18a}. For the remaining four galaxies in the SOFIA sample where no HAWC+ observations are available, we instead use the SFRs reported in \citet{green14} to estimate TIR and the calibration in equation (3) of \citet{cluver17}:

\begin{equation}
    \mathrm{SFR\;[M_{\odot}\,yr^{-1}]} = 2.8\,\times\,10^{-44}\,L_{TIR} \; [\mathrm{erg\,s^{-1}}]
\end{equation}

\noindent which is derived from \textsc{Starburst99} for solar metallicity, continuous star formation over 100~Myr, a Kroupa IMF, and assumes that the ultraviolet (UV) component of stellar emission is completely absorbed and re-radiated in the infrared \citep[see also][]{calzetti13}. 

To determine the \CII{} luminosities, we produce integrated intensity maps from the FIFI-LS observations (see Figure \ref{fig:fifi_obs} in Appendix \ref{app:sofia}) and take the peak value within a beam located at the position of each galaxy. In  the right panel of Figure \ref{fig:all_sed}, we present the \CII{}/TIR as a function of TIR measured in this sample of DYNAMO galaxies. The magenta squares represent galaxies for which SEDs were used to derive TIR, while the teal diamonds represent the galaxies for which the SFRs were used instead. In both cases, we show error bars where the errors on the \CII{} and TIR luminosities have been propagated into the ratio. The two approaches to estimating the TIR luminosities yield consistent results. To illustrate this, we join with black dashed lines the data points for which we have SEDs and SFRs. When we compare DYNAMO to existing measurements in different types of galaxies \citep{herrera-camus18a}, we find that DYNAMO galaxies do not exhibit a \CII{}-deficit. \citet{herrera-camus18a} shows that at a fixed IR luminosity, the \CII{}/FIR ratio decreases as galaxies become more compact, and \citet{lutz16} shows that the line-to-FIR ratios form a much tighter relation with FIR surface brightness than luminosity. To investigate this, \citet{herrera-camus18b} construct two toy models with the PDR toolbox \citep{kaufman06}: one where OB stars are closely associated with molecular gas clouds, and another where OB stars and clouds are randomly distributed. They find that as galaxies become more compact, a combination of effects give rise to the \CII{}-deficit. These include a reduction of the photo-electric heating efficiency, an increase in the ionization parameter, and as the interstellar radiation field increases, the \CII{} line saturates and becomes nearly independent of the far-UV flux. Although DYNAMO galaxies generally lie above the star-forming main sequence at $z \sim 0.1$, their star formation is distributed throughout their disks within numerous star-forming clumps, rather than being confined to a compact region. Their low dust temperatures and lack of a \CII{}-deficit is consistent with this morphology.

\section{Discussion} \label{sec:discussion}
\begin{figure}
    \centering
    \includegraphics[width=\columnwidth]{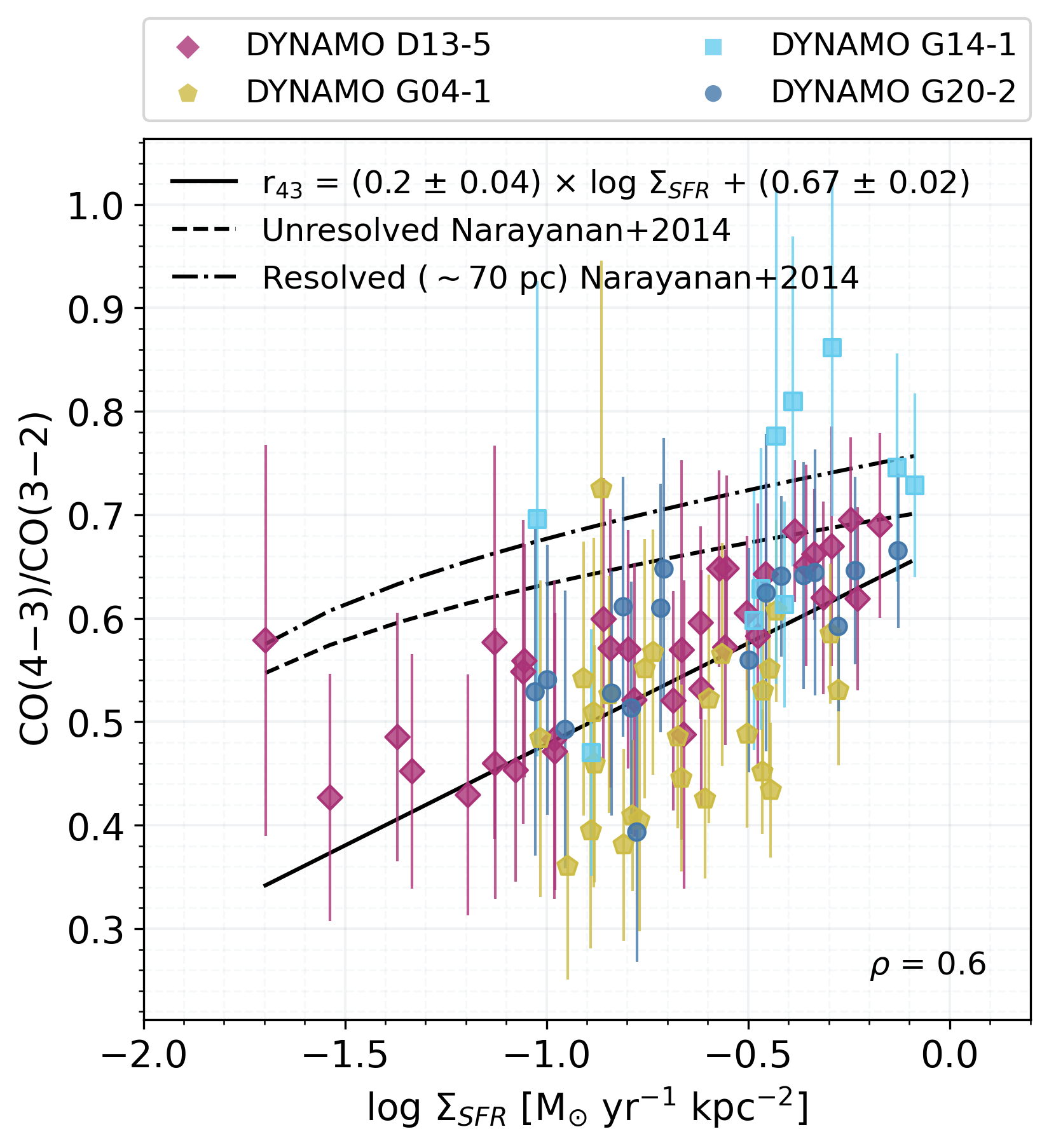}
    \caption{\jupfour{}/\jupthree{} line ratio as a function of SFR surface density, measured in beam-sized regions across the disk of each galaxy, indicated by the color and symbol coding in the legend. We present this data for galaxies where observations of both CO transitions and H$\alpha$ exist. Despite the large uncertainties, there is an indication of an increasing line ratio with increasing SFR surface density trend. This is parameterized by the Spearman's Rank Order correlation coefficient of $\rho = 0.6$, suggesting a moderate positive correlation between these two quantities. We present a linear fit to our measurements (black solid line) and for comparison, include the predicted trend using the unresolved relation between CO intensity and \sfrsd\ of \citet{narayanan14} (black dashed line), and their $70$~pc resolved relations (black dash-dotted line)}.
    \label{fig:sfrsd_lr}
\end{figure}

The $\sim 1-2$~kpc-scale ALMA observations allow us to investigate how the line ratios we measure are affected by the surface density of star formation. We expect that the \jupfour\ transition will be more highly excited in regions of higher \sfrsd{}, because these regions will have larger UV radiation fields and thus warmer dust temperatures \citep{narayanan14}. To test this, we compare our resolved line ratio measurements to the \sfrsd{} measurements we make in the same beam-sized apertures. Figure \ref{fig:sfrsd_lr} shows the \jupfour{}/\jupthree{} line ratio as a function of \sfrsd{} for four galaxies for which all necessary observations are available, as indicated by the legend. For each galaxy, we plot the set of resolved beam-sized measurements as previously described. Though the line ratio uncertainties are large, there is a moderate positive correlation between the line ratios and \sfrsd{} measurements, indicating that in this sample of DYNAMO galaxies, higher \sfrsd{} regions are indeed correlated with higher line ratios. We perform a Spearman Rank Order correlation and find a coefficient of $\rho = 0.6$. The correlation between resolved measurements within a single galaxy is stronger for DYNAMO D13-5 and G20-2 ($\rho = 0.8, 0.7$ respectively), and weakest for DYNAMO G04-1 ($\rho = 0.4$), while for G14-1 it is $\rho = 0.6$. In addition, we perform a linear fit to our observed line ratio$-$\sfrsd\ relation using \texttt{scipy.curve\_fit}, which performs a non-linear least squares analysis with errors on the $y-$data as a parameter, and show the results with the black solid line. The black dashed line corresponds to the parameterization of CO line emission intensity as a function of \sfrsd{}, derived by \citet{narayanan14} (their equation 19): 

\begin{equation}
    \frac{I_{ij}}{I_{1-0}} = A \times [\mathrm{log}_{10}(\Sigma_{\mathrm{SFR}}) - \chi]^{B} + C
    \label{eq:sfrsd_lr}
\end{equation}

\noindent where $I_{ij}$ is the intensity of the CO($i-j$) transition, $A$, $B$, and $C$ are fit parameters, and $\chi = -1.85$ (an offset introduced to produce only real values of $I_{ij}/I_{1-0}$). \citet{narayanan14} calculate the CO SLED of high$-z$ star-forming galaxies from CO intensities that are modeled at $\sim 70$~pc resolution. For real observations with coarser beams, such as in our case, the resolved line ratio$-$\sfrsd{} parameterization is not an appropriate comparison. Therefore, \citet{narayanan14} determine the luminosity-weighted emitting area for each CO transition and scale the resolved line intensities, and then refit the line ratio$-$\sfrsd{} relation. Because our observations probe $\sim 1-2$~kpc scales, this is primarily what we compare to here. However, we show comparisons to the resolved parameterization for completeness We adopt values for $A$, $B$, and $C$ for unresolved observations from their Table 3 for \jupthree{} and \jupfour{}, and substitute in our measured values of \sfrsd{}. Finally, we take the ratio of the two equations and divide by $J_{u}^{2}/J_{l}^{2} = 4^{2}/3^{2}$ to convert from Jy to K and produce the dashed black line in Figure \ref{fig:sfrsd_lr}. We repeat the same procedure for the resolved galaxy observations parameterization from their Table 2 and plot this as the black dashed-dotted line in Figure \ref{fig:sfrsd_lr}.

Overall, both model parameterizations under-predict the steepness of the \jupfour{}/\jupthree{}$-$\sfrsd{} relation that our observations suggest, and over-predict the line ratio across the entire range of \sfrsd{} values that our observations probe. Similarly, \citet{boogaard20} find that the unresolved models also over-predict their \jupfour{}/\juptwo{} measurements (see their Figure 13), while providing a better match to their \jupfive{}/\juptwo{} values. \citet{sharon19}, who present $\sim 2$~kpc resolution \jupone{} and \jupthree{} observations of a lensed galaxy at $z = 2.26$, also find that the \citet{narayanan14} models do not reproduce their observations; however, they do not attribute much meaning to this difference due to the limited \sfrsd{} values probed by a single galaxy. In contrast, \citet{valentino20} find qualitative agreement between the unresolved \citet{narayanan14} model and their \jupfive{}/\juptwo{} observations in $z = 1.1-1.7$ IR-selected galaxies on and above the main sequence of star formation. 

It is possible that because these models do not explicitly model gas-rich clumpy disks like DYNAMO and high-redshift star-forming galaxies, that their properties are not completely captured in the early-phase snapshots of the model disks and model mergers \citep{narayanan14}. It is also possible that a model which characterizes global CO excitation properties for an average \sfrsd{} may not be well-suited to investigate the internal variations within a single galaxy. To test this, we convolve our H$\alpha$ maps to the \jupone{} beam sizes ($\sim 5-10$\arcsec{}) of \citet{fisher19} and measure the global \sfrsd{} of each galaxy for which data are available. We then use \ref{eq:sfrsd_lr} and the unresolved parameters of \citet{narayanan14} to predict $R_{31}$ and $R_{41}$. We list these predictions in the last two columns of Table \ref{tab:lrs_co10}. We find that overall, the \citet{narayanan14} models give better predictions of our global $R_{31}$ and $R_{41}$ measurements than our kpc-scale $R_{43}$ measurements, which may indicate that the unresolved models do not capture the kpc-scale variation in CO excitation. Using hydrodynamical simulations, \citet{bournaud15} study the CO SLEDs of high-redshift galaxies (as well as spirals and mergers), and investigate the contribution of giant clumps to the global CO SLED. They derive CO SLEDs for clumps and the inter-clump gas and show that there is a considerable difference in the CO excitation (see their Figures 3 and 4). This may indicate a need for models that specifically relate CO excitation, measured at various physical scales, in gas-rich clumpy disks to observable quantities such as \sfrsd{}.



\section{Conclusions} \label{sec:conclusion}
In this work, we have combined $\sim 1-2$~kpc scale ALMA observations of \jupthree{} and \jupfour{} with HST observations of H$\alpha$, to study the \jupfour{}/\jupthree{} line ratio and its dependence on \sfrsd{}. We have combined this with SOFIA HAWC+ and FIFI-LS observations of \CII{} which provide additional measurements of the ISM gas physical conditions. We summarize our findings here: \\

\noindent 1. DYNAMO galaxies have typical \jupfour{}/\jupthree{} line ratios of $R_{43} = 0.54${\raisebox{0.5ex}{\tiny$^{+0.16}_{-0.15}$}}, which is most consistent with samples of star forming $\sim 1-2$ main-sequence galaxies \citep[e.g.,][]{daddi15,boogaard20,henriquez-brocal22}. \\

\noindent 2. Likewise, the global \jupthree{}/\jupone{} and \jupfour{}/\jupone{} measurements in DYNAMO are higher than global measurements of nearby star-forming galaxies \citep{leroy22} and are more consistent with the measurements of $z \sim 1-2$ star-forming galaxies \citep[see e.g.,][]{daddi15,dessauges-zavadsky15,birkin21,harrington21}. \\

\noindent 3. The DYNAMO SEDs derived from SOFIA HAWC+ suggest cooler dust temperatures than those observed in local starburst galaxies and U/LIRGs. This suggests that while DYNAMO galaxies are strongly star forming, their star formation must be distributed rather than very compact. This is consistent with the picture we obtain from the \jupfour/\jupthree{} line ratio measurements and the clumpy morphology of these systems. \\

\noindent 4. The DYNAMO \jupfour/\jupthree\ line ratios are positively correlated with the \sfrsd\ measurements, with a Spearman Rank Order correlation coefficient of $\rho = 0.6$. Our best fit relation between the \jupfour/\jupthree{} line ratio and \sfrsd\ is $R_{43} = (0.2 \pm 0.04)\, \times\, \mathrm{log}\, \Sigma_{\mathrm{SFR}}\, +\, (0.67 \pm 0.02)$. This relation suggests a steeper relation than predicted by the parameterization of \citet{narayanan14}, which also over-predicts the line ratio over the whole range of \sfrsd{} values probed by observations. It is possible that this is consistent with the low dust temperatures of DYNAMO galaxies. However, \citet{sharon19}, who also study $\sim$~kpc scale line ratios in a high-redshift lensed galaxy, also find a discrepancy between the models and observations. This may indicate that models that investigate CO emission variations with internal galaxy properties for gas-rich clumpy disks, are required. 

\acknowledgments
We thank the anonymous referee for comments and suggestions that have greatly improved this work. This paper makes use of the following ALMA data: ADS/JAO.ALMA\#2017.1.00239.S. and ADS/JAO/ALMA\#2019.1.00447.S. ALMA is a partnership of ESO (representing its member states), NSF (USA) and NINS (Japan), together with NRC (Canada), MOST and ASIAA (Taiwan), and KASI (Republic of Korea), in cooperation with the Republic of Chile. The Joint ALMA Observatory is operated by ESO, AUI/NRAO and NAOJ. The National Radio Astronomy Observatory is a facility of the National Science Foundation operated under cooperative agreement by Associated Universities, Inc. Some of the data presented in this paper were obtained from the Mikulski Archive for Space Telescopes (MAST) at the Space Telescope Science Institute. The specific observations analyzed can be accessed via \dataset[10.17909/faa7-sw34]{https://doi.org/10.17909/faa7-sw34}. Based in part on observations made with the NASA/DLR Stratospheric Observatory for Infrared Astronomy (SOFIA). SOFIA is jointly operated by the Universities Space Research Association, Inc. (USRA), under NASA contract NNA17BF53C, and the Deutsches SOFIA Institut (DSI) under DLR contract 50 OK 0901 to the University of Stuttgart. Financial support for this work was provided by NASA through award \#SOFIA-080238 issued by USRA. L.L. and A.D.B. acknowledges support from USRA SOFIA-080238 and NASA HSTGO15069002A, and NSF-AST2108140. R.C.L. acknowledges support from a NSF Astronomy and Astrophysics Postdoctoral Fellowship under award AST-2102625. D.O. is a recipient of an Australian Research Council Future Fellowship (FT190100083) funded by the Australian Government. R.H.-C. thanks the Max Planck Society for support under the Partner Group project "The Baryon Cycle in Galaxies" between the Max Planck for Extraterrestrial Physics and the Universidad de Concepción. R.H-C also acknowledge financial support from Millenium Nucleus NCN19058 (TITANs) and support by the ANID BASAL projects ACE210002 and FB210003. This research made use of Photutils, an Astropy package for detection and photometry of astronomical sources \citep{larry_bradley_2021_5525286}.

%

\vspace{5mm}
\facilities{ALMA, HST(WFC), SOFIA(FIFI-LS, HAWC+)}


\software{aplpy \citep{aplpy2019},
          astropy \citep{astropy13,astropy18},
          \casa\ \citep{mcmullin07},
          numpy \citep{2020NumPy-Array},
          Photutils \citep{larry_bradley_2021_5525286},
          reproject \citep{robitaille_thomas_2018_1162674}, 
          spectral-cube \citep{adam_ginsburg_2019_3558614} 
          }



\newpage
\appendix

\section{SOFIA Observations} \label{app:sofia}
\begin{figure*}
    \centering
    \includegraphics[height=5.1cm]{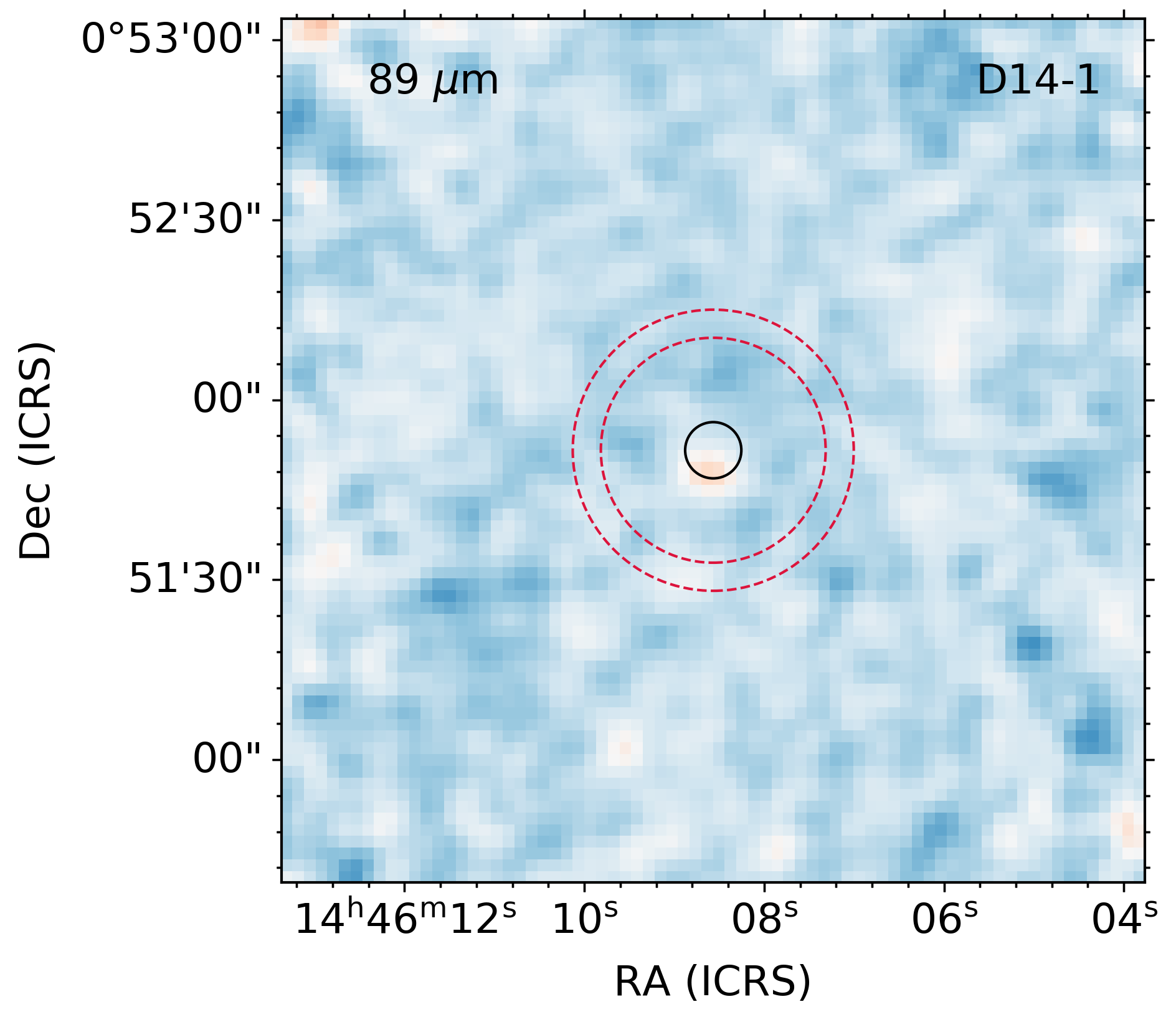}
    \hspace{0.28cm}
    \includegraphics[height=5.1cm]{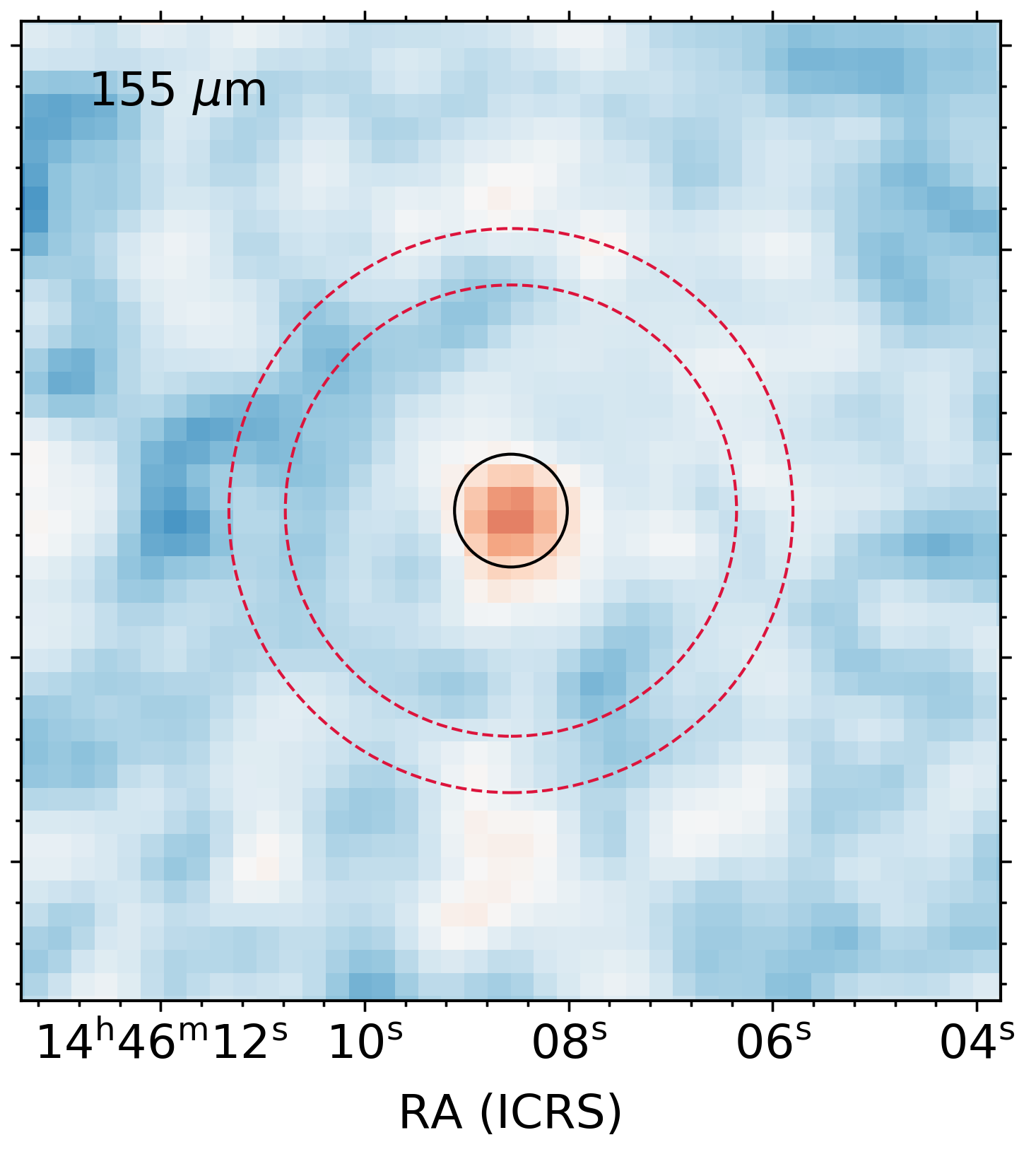}
    \hspace{0.28cm}
    \includegraphics[height=5.1cm]{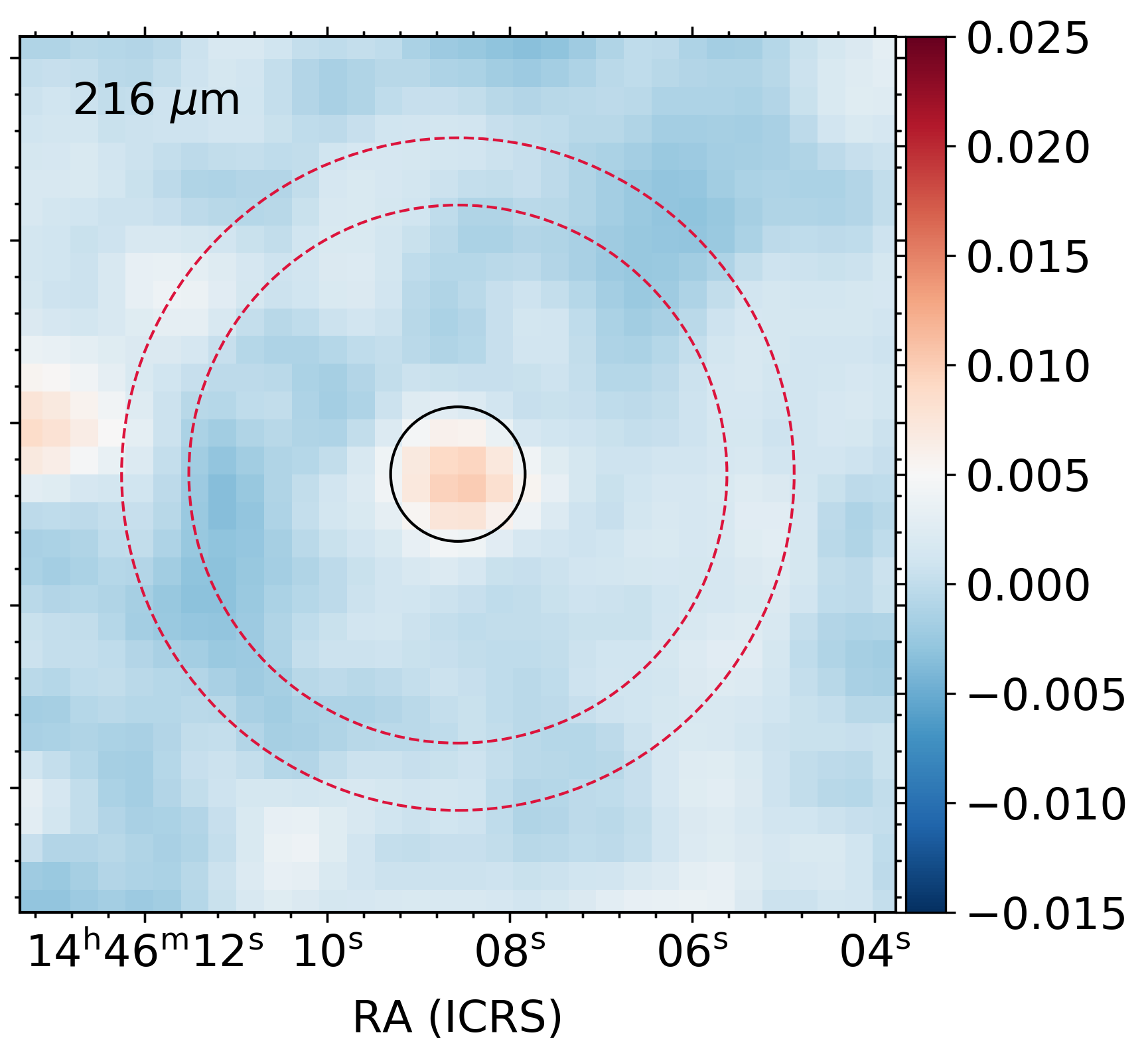}
    
    \includegraphics[height=5.1cm]{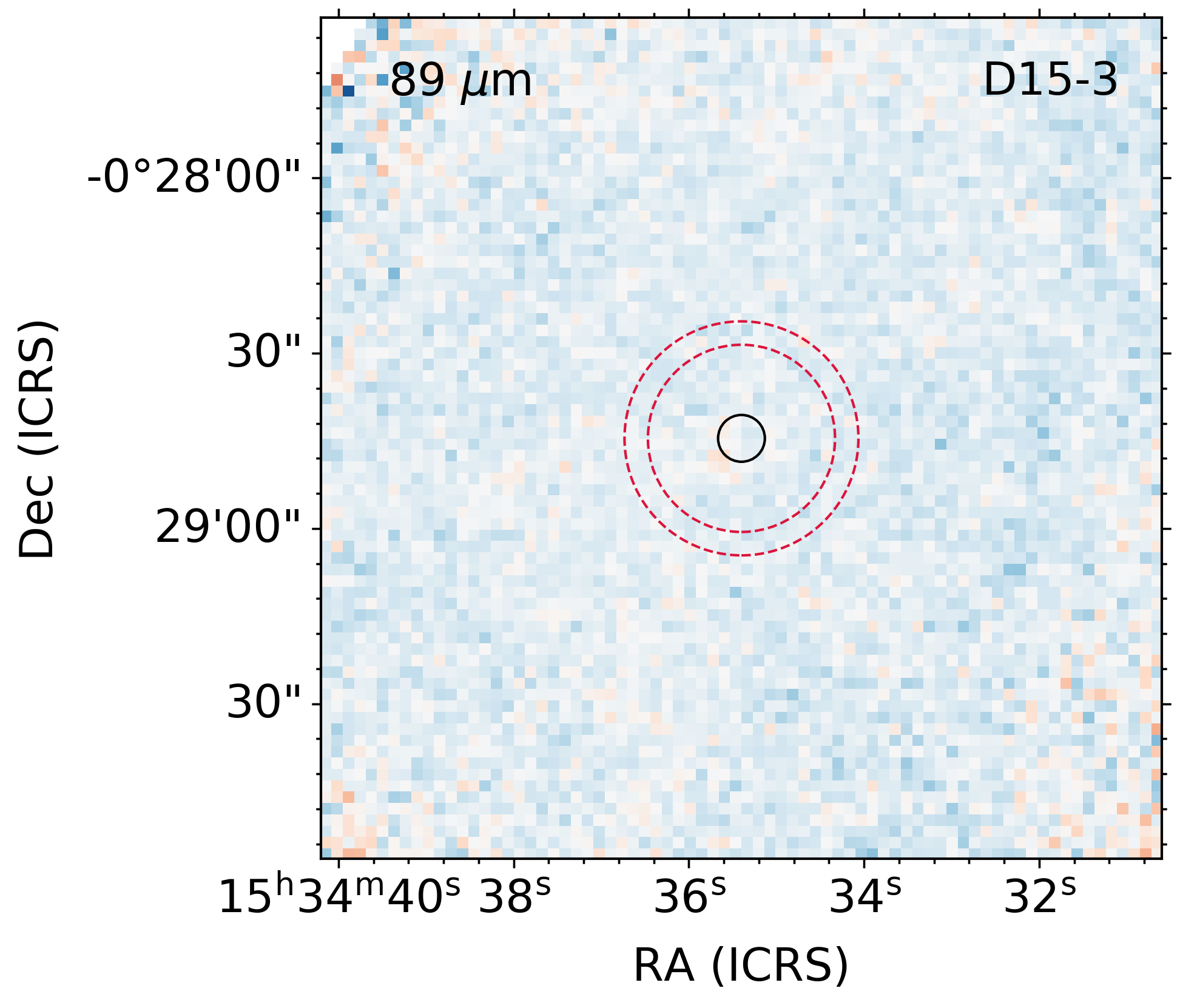}
    \includegraphics[height=5.1cm]{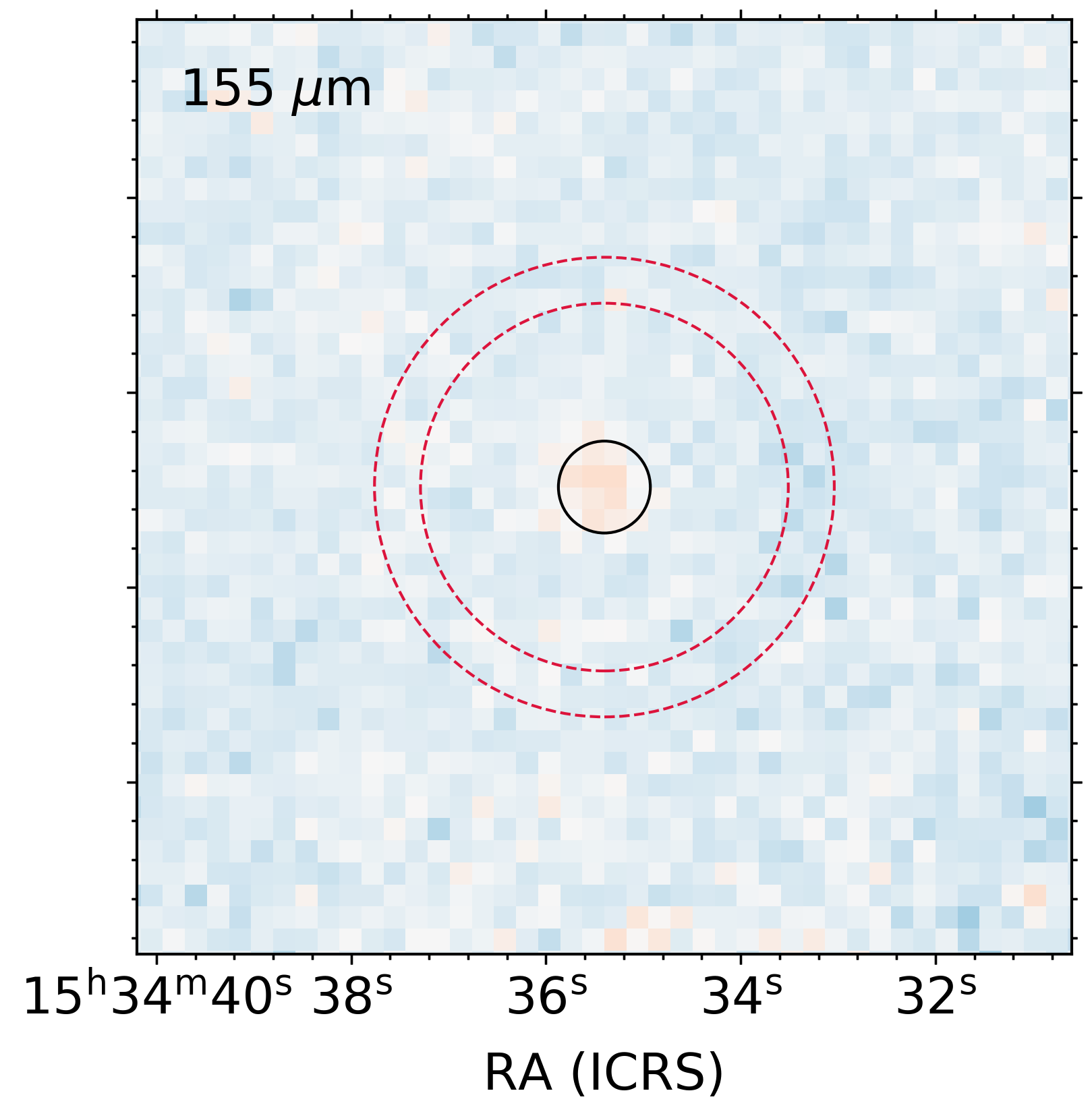}
    \includegraphics[height=5.1cm]{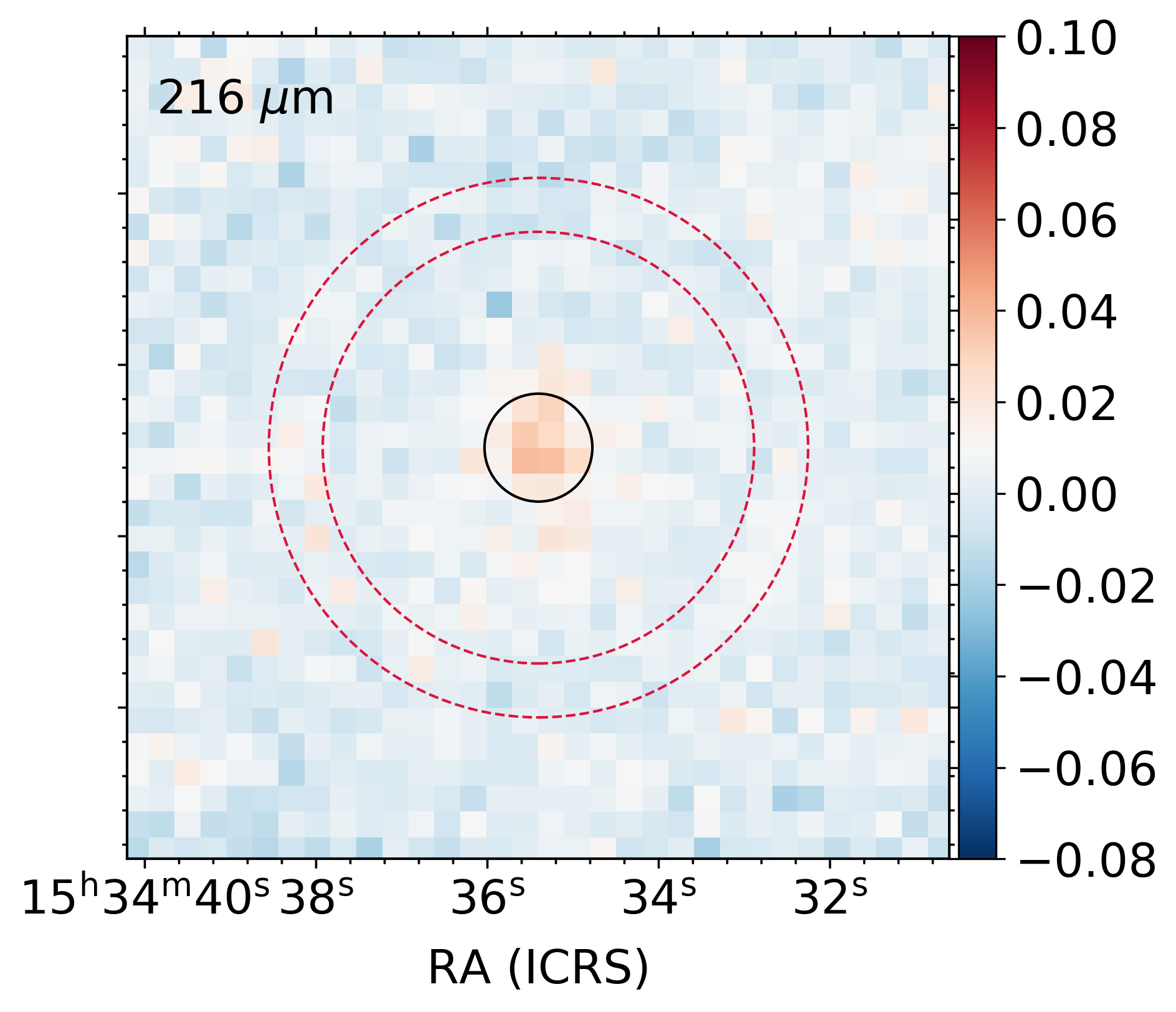}
    
    \includegraphics[height=5.1cm]{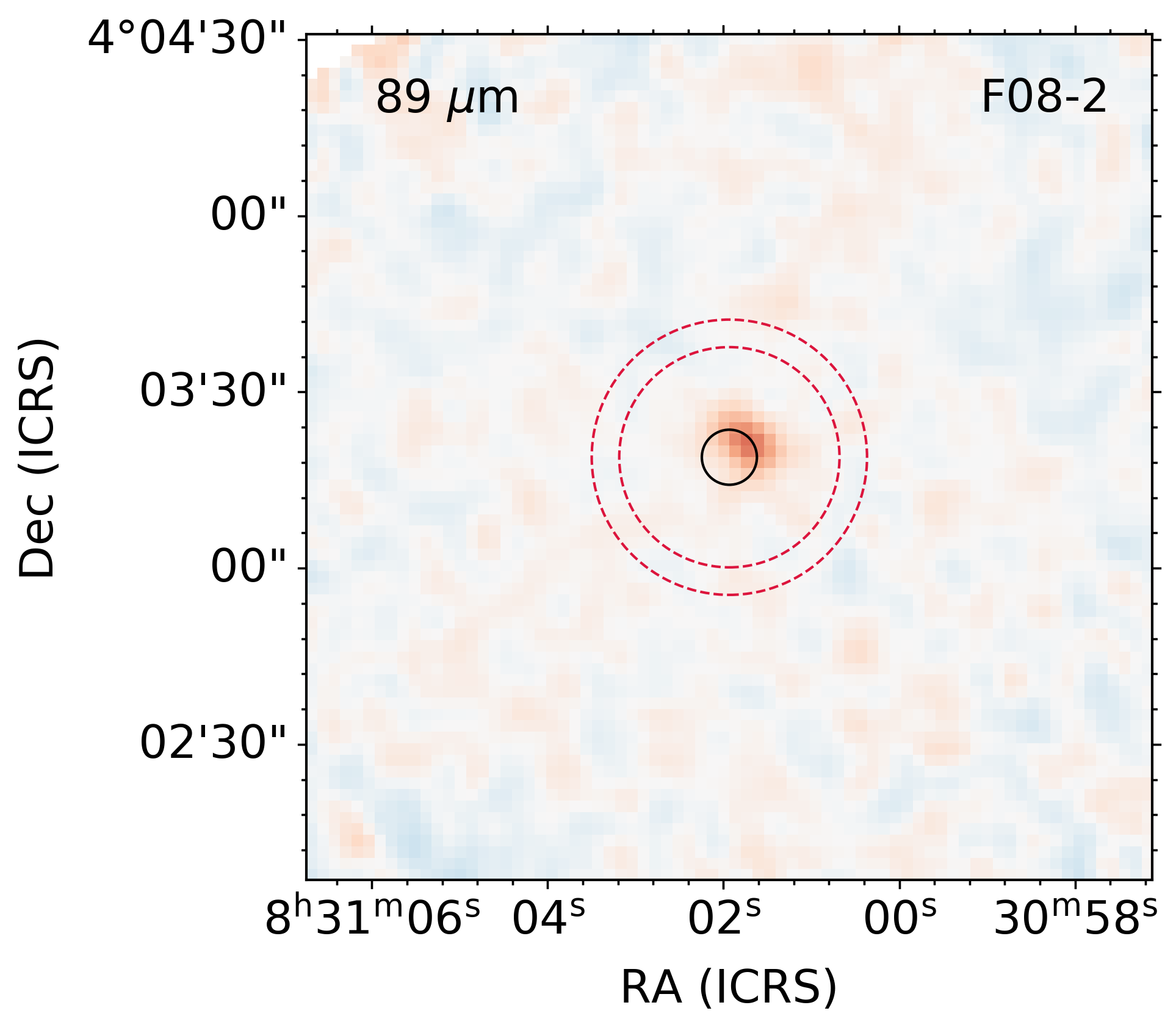}
    \hspace{0.18cm}
    \includegraphics[height=5.1cm]{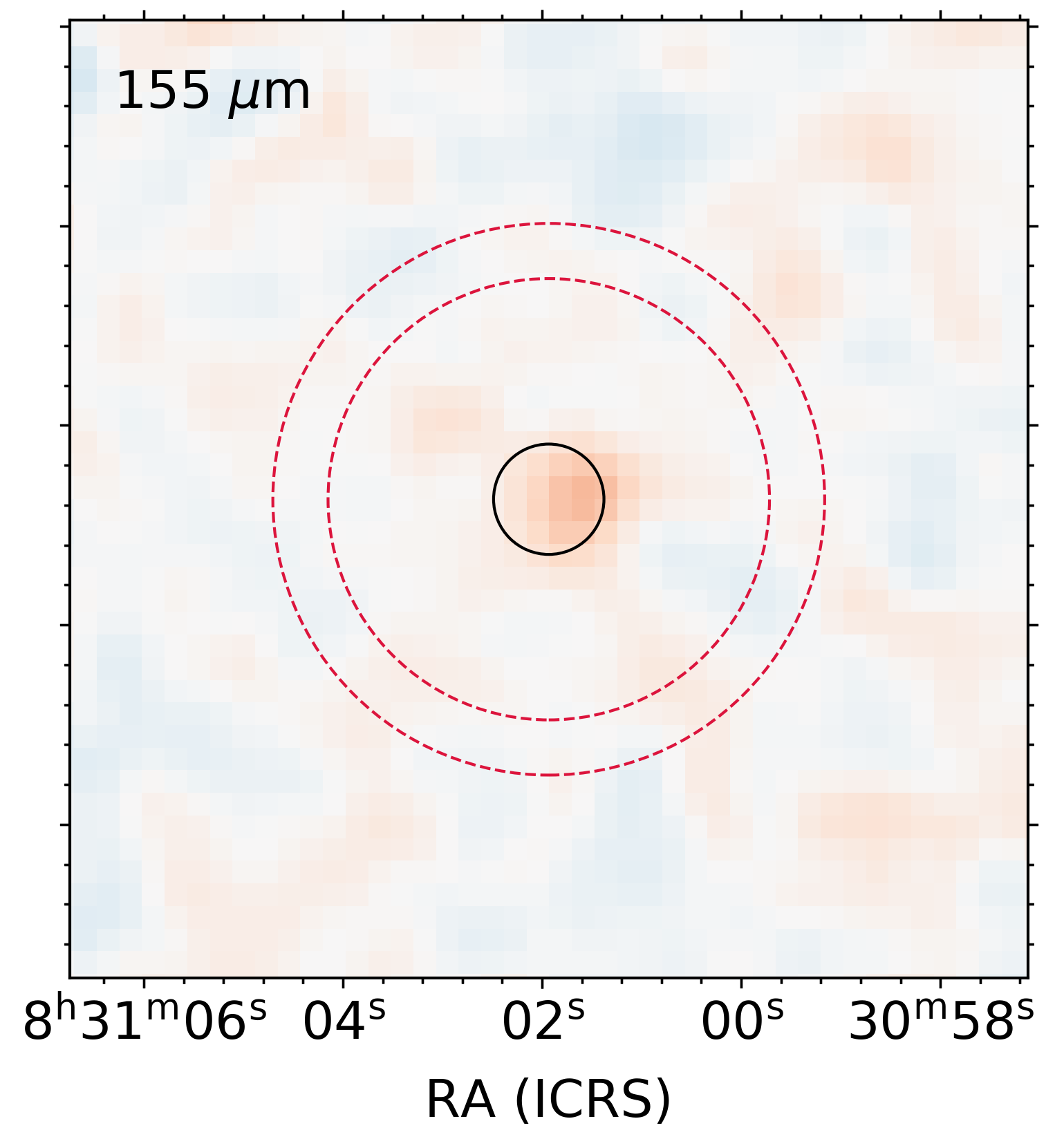}
    \hspace{0.18cm}
    \includegraphics[height=5.1cm]{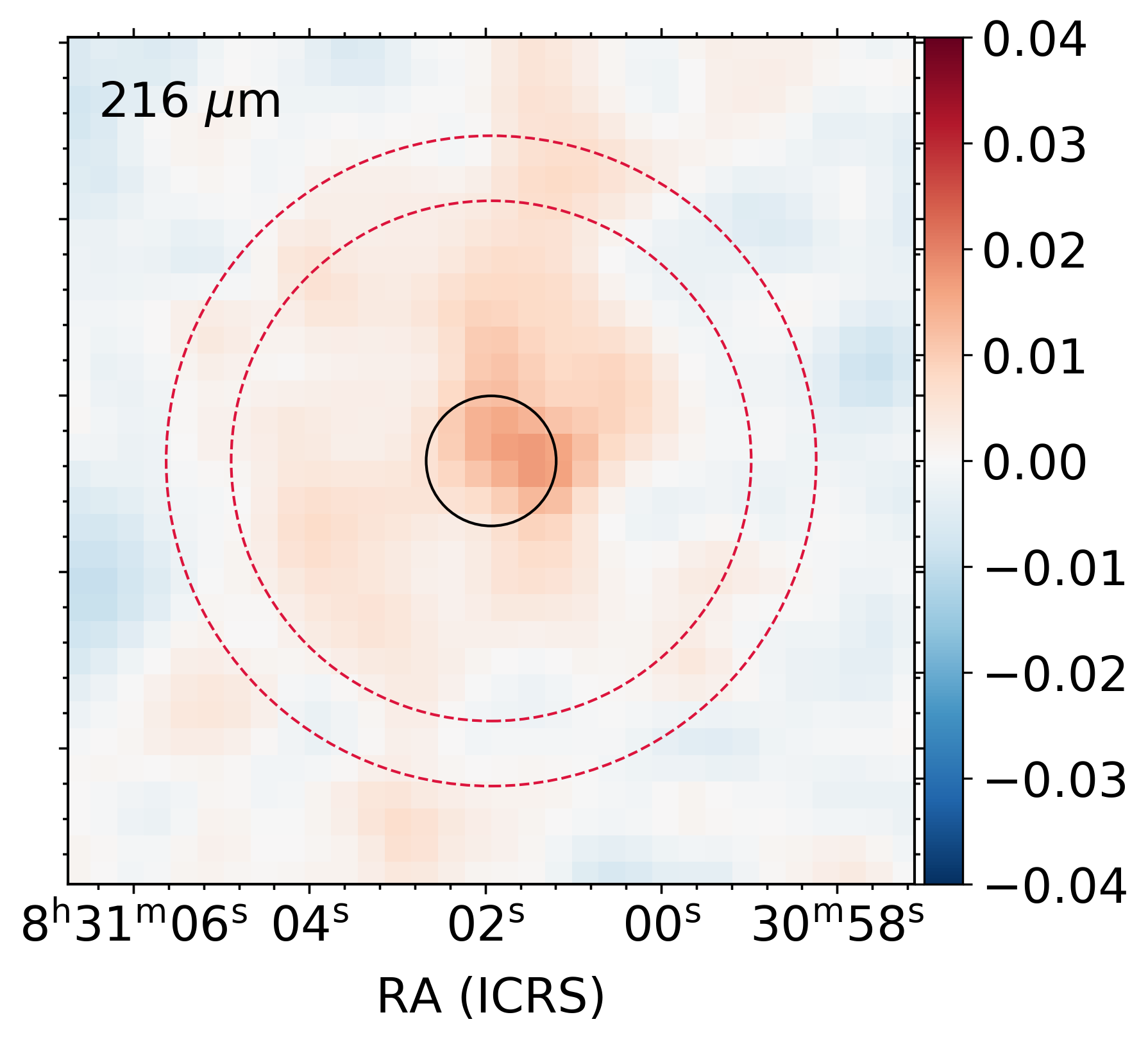}
    
    \includegraphics[height=5.1cm]{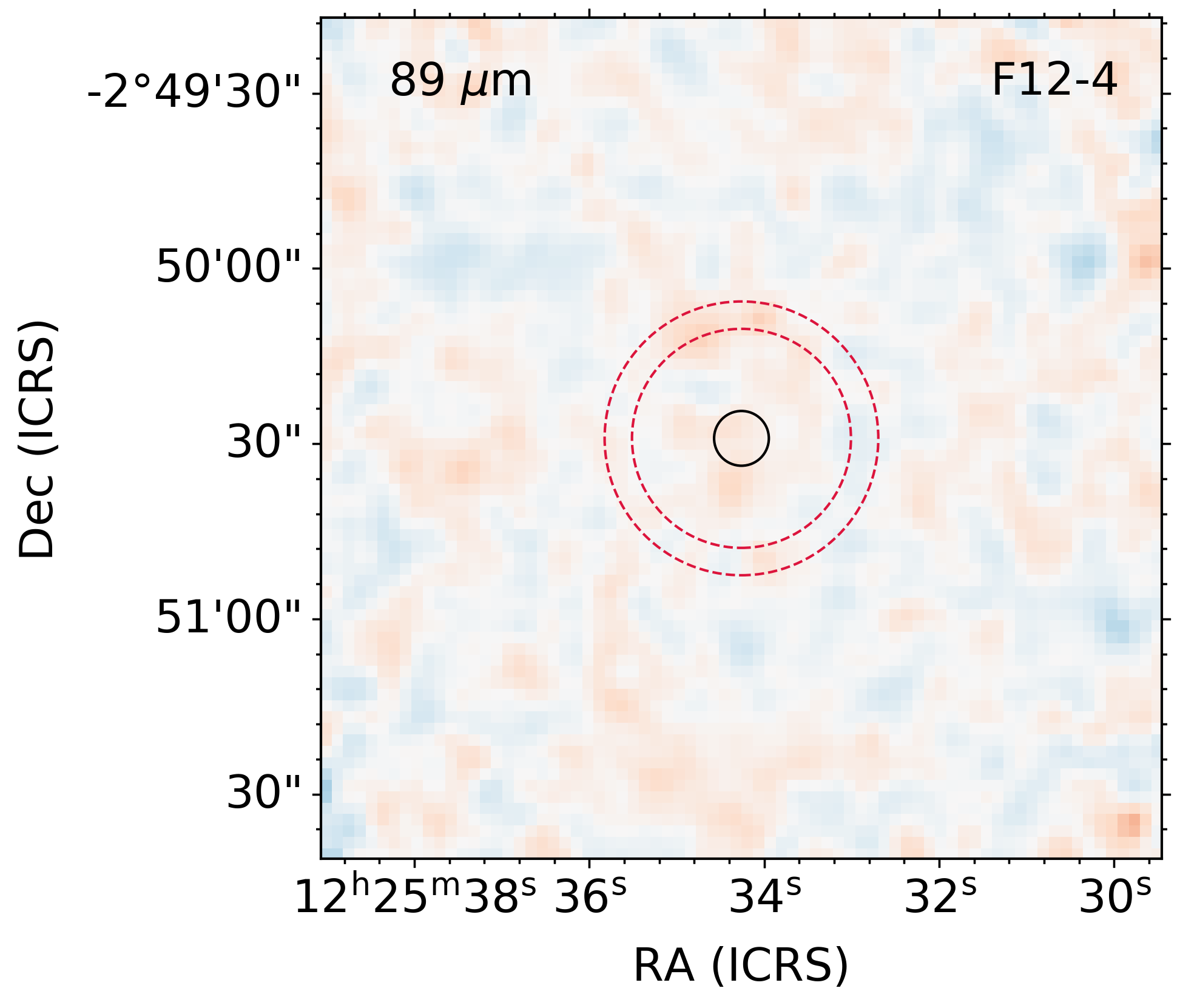}
    \hspace{0.18cm}
    \includegraphics[height=5.1cm]{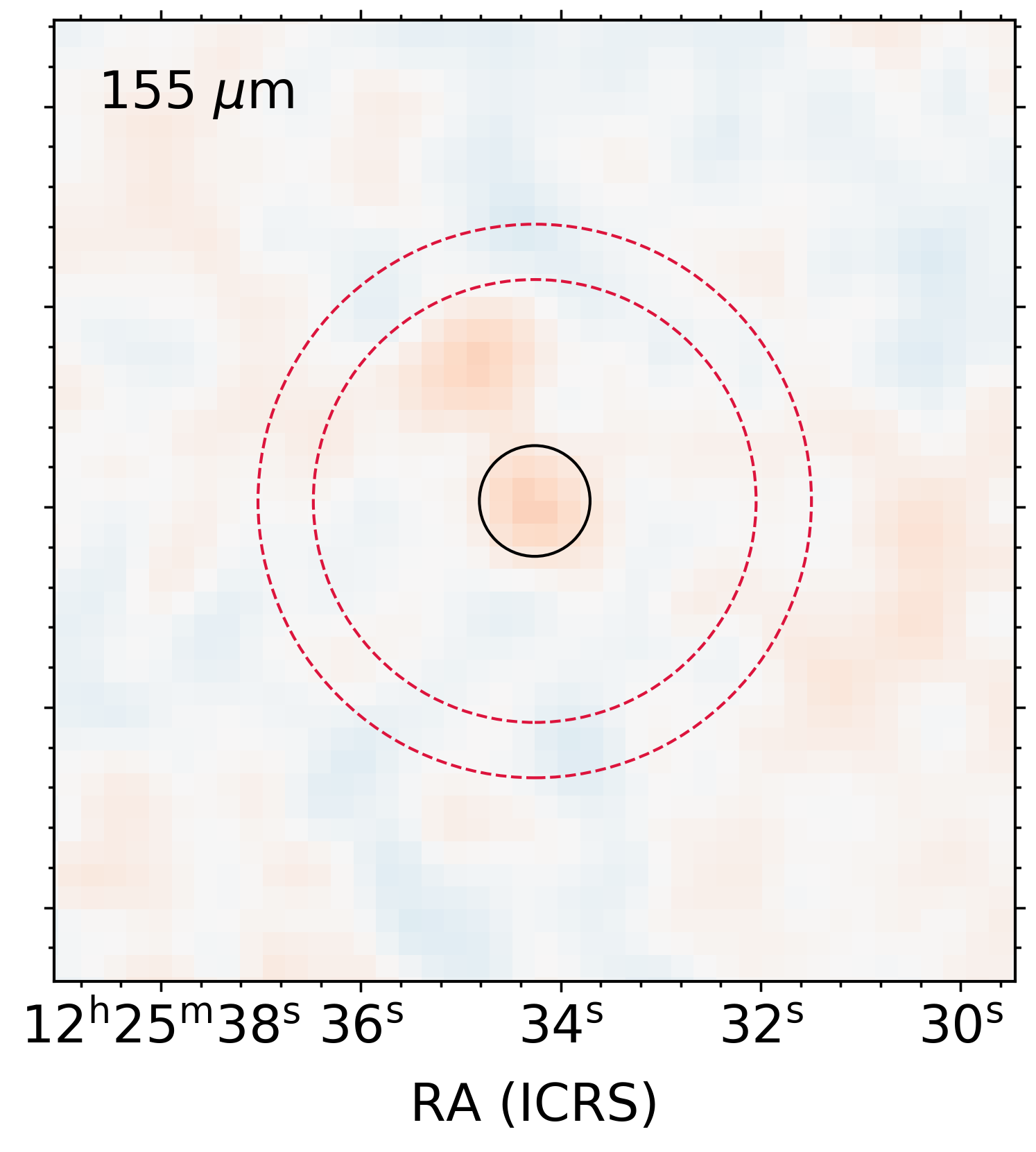}
    \hspace{0.18cm}
    \includegraphics[height=5.1cm]{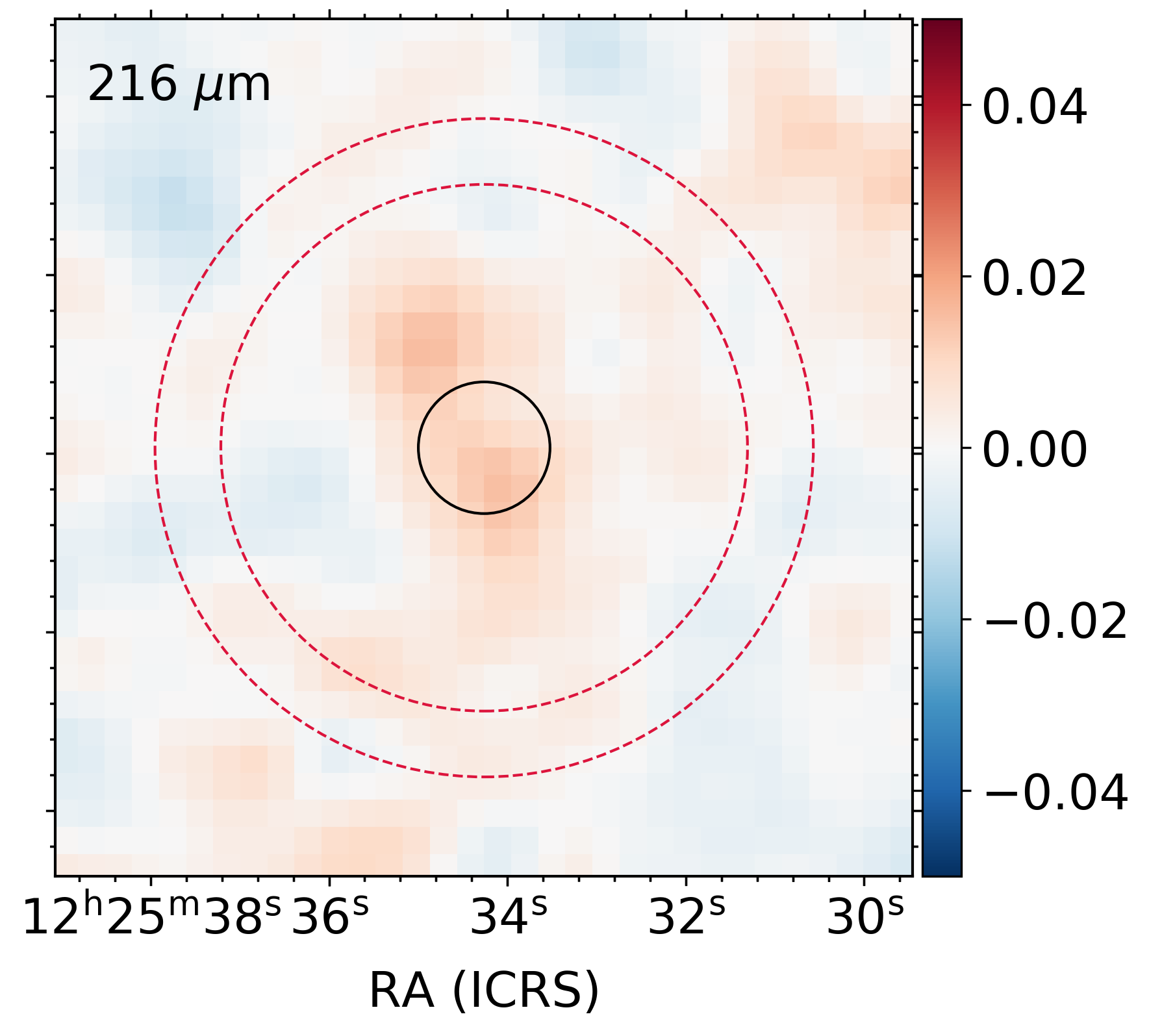}
    \caption{SOFIA HAWC+ observations of DYNAMO galaxies: 89~\micron{} (left), 155~\micron{} (middle), and 216~\micron{} (right) in units of Jy\,pixel$^{-1}$. The black circles are centered on the position of the DYNAMO galaxy observed (indicated in the top right corner of the left-most panels), and their size corresponds to the angular resolution of each band (wavelength is indicated in the top left corner of each panel). The crimson dashed circles define the background annulus. DYNAMO D15-3 (second row) is the only galaxy that overlaps with the ALMA sample.}
    \label{fig:hawc_obs}
\end{figure*}

\begin{figure*}
    \centering
    \includegraphics[height=4.5cm]{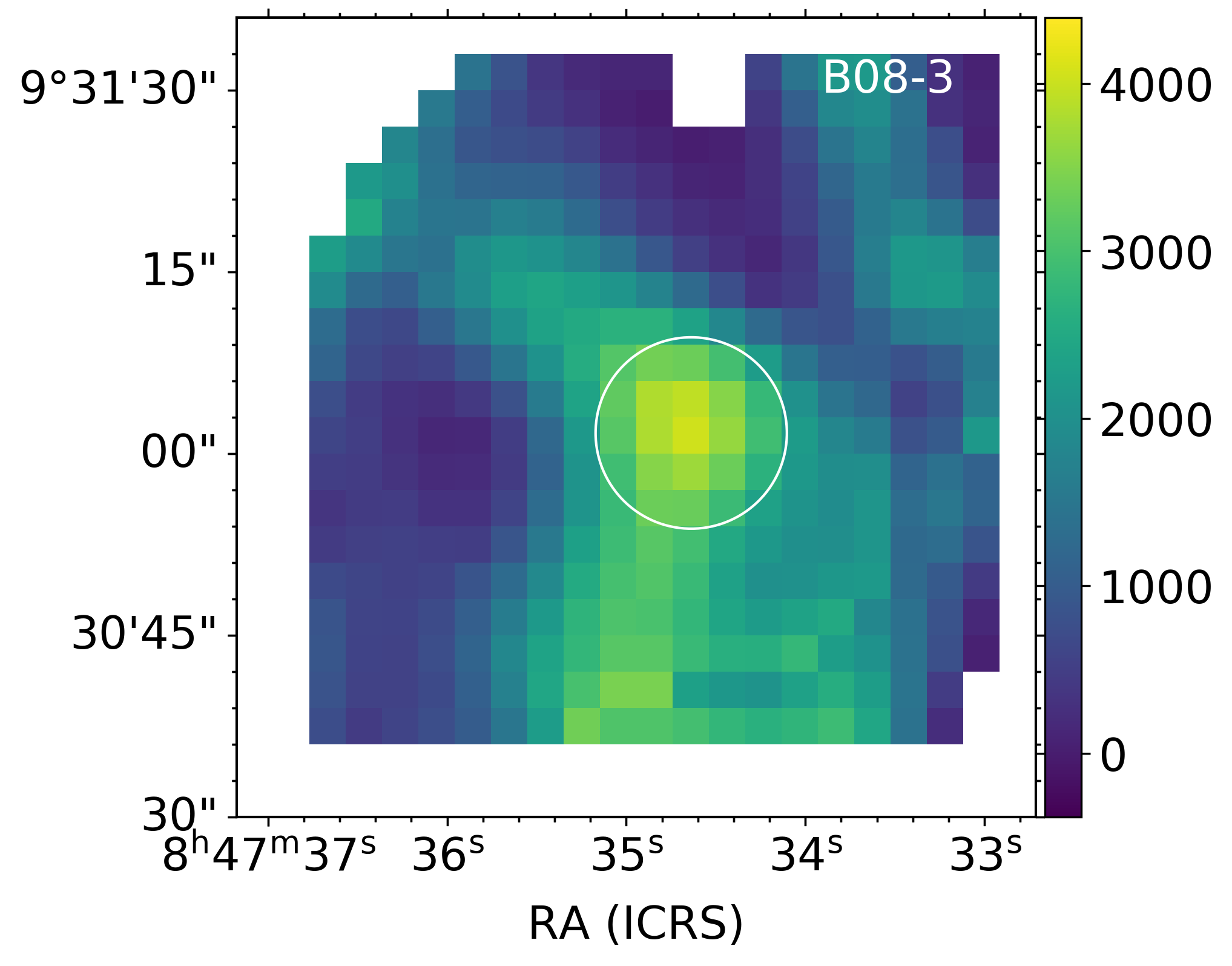}
    \includegraphics[height=4.5cm]{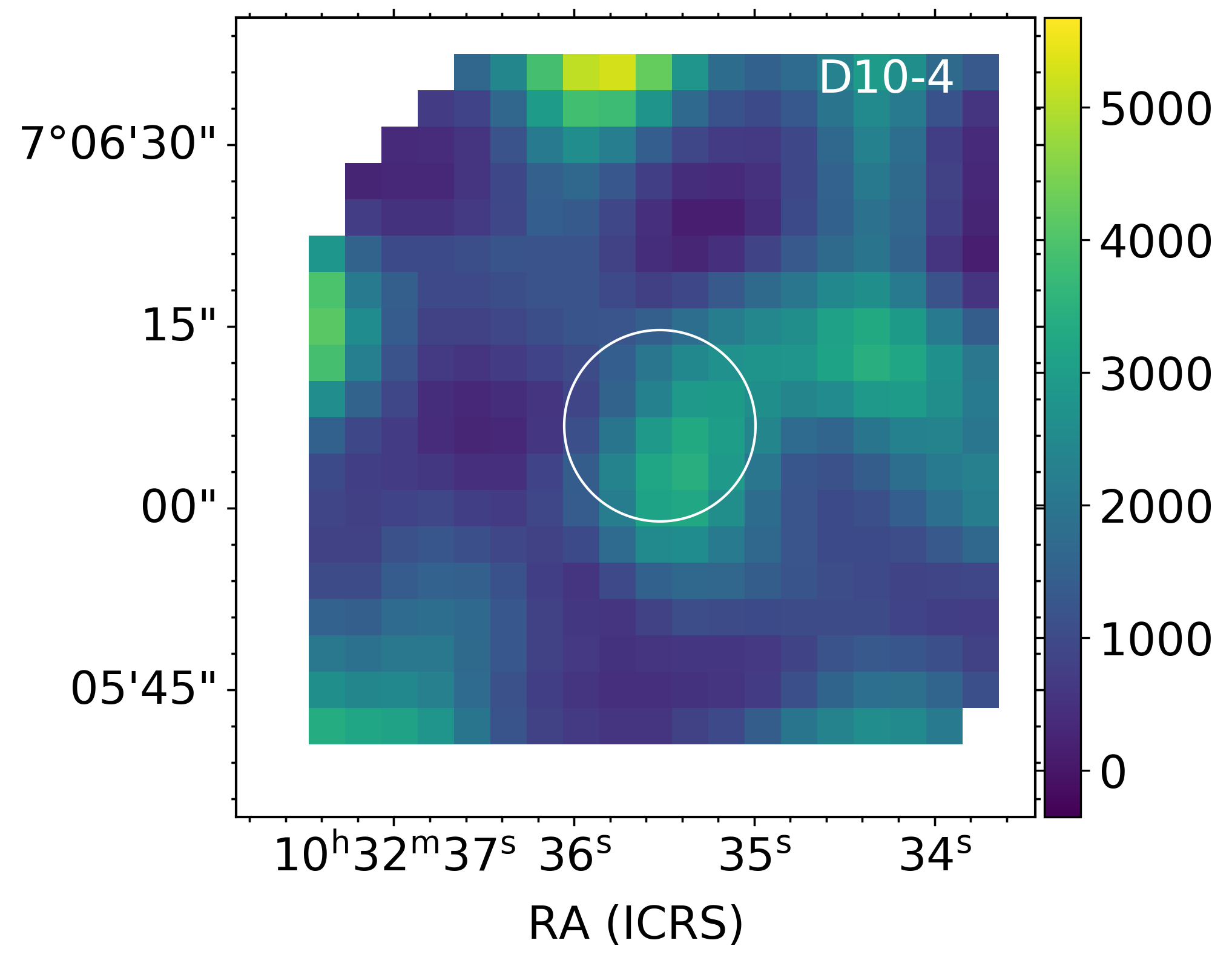}
    \includegraphics[height=4.5cm]{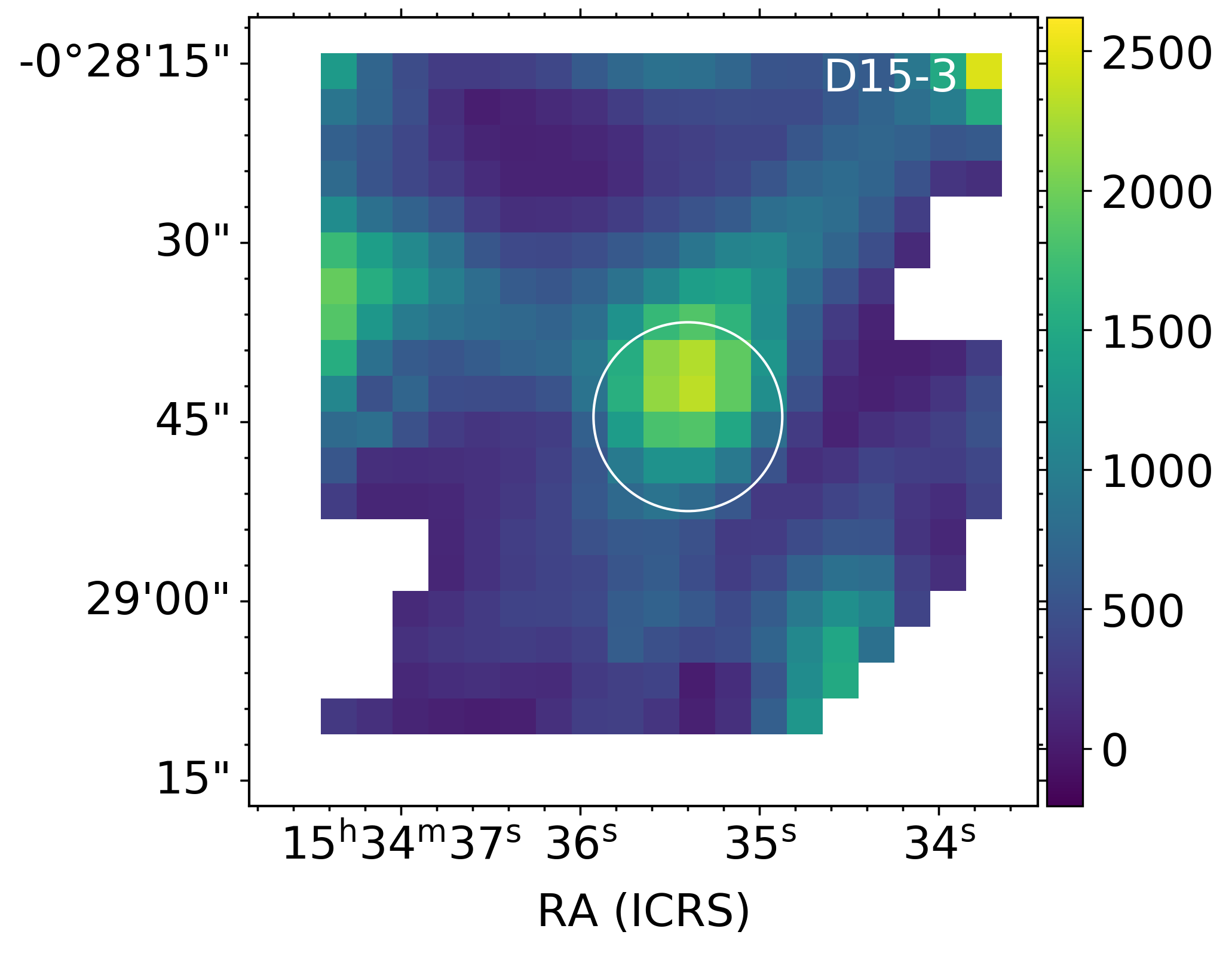}
    
    \includegraphics[height=4.5cm]{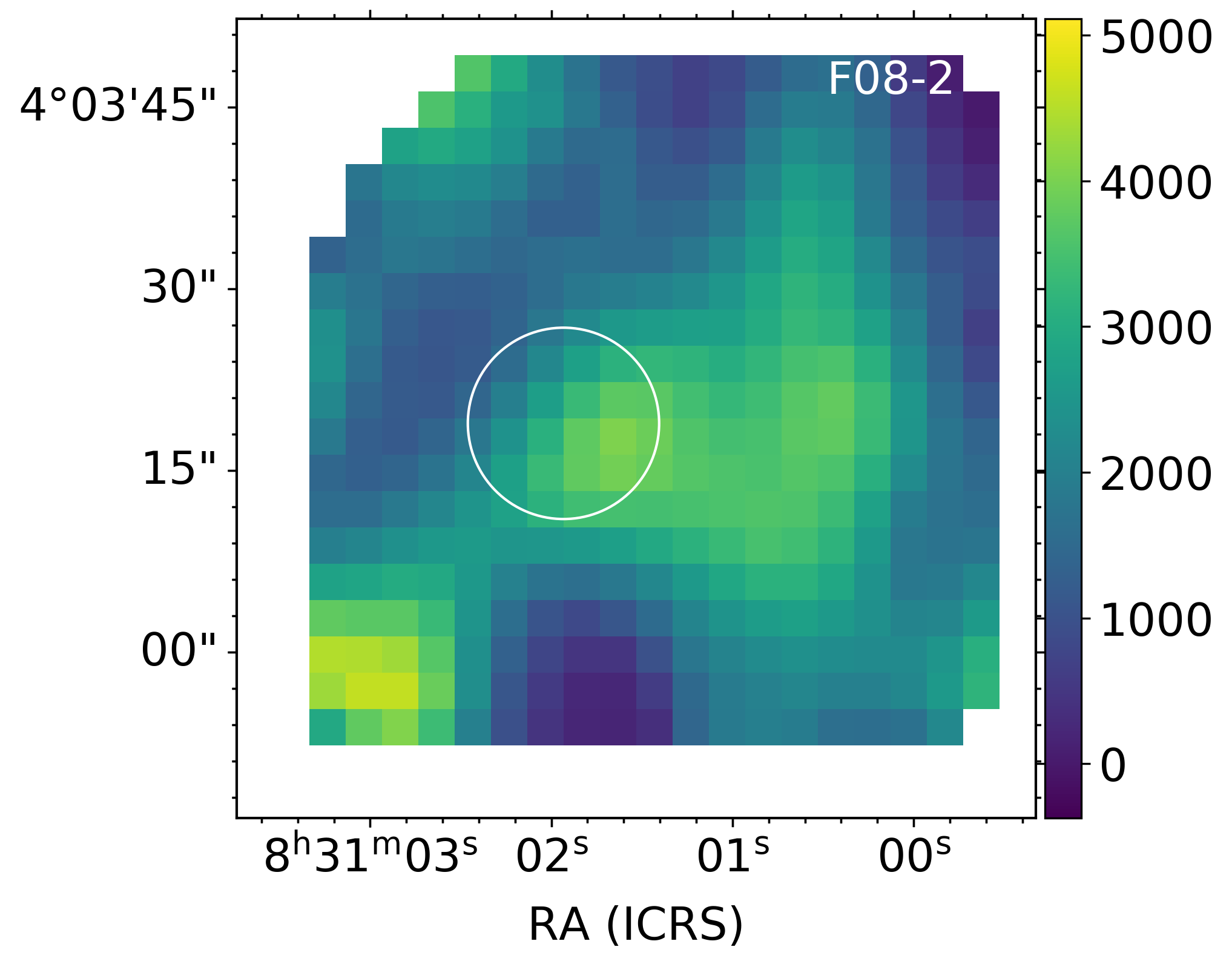}
    \includegraphics[height=4.5cm]{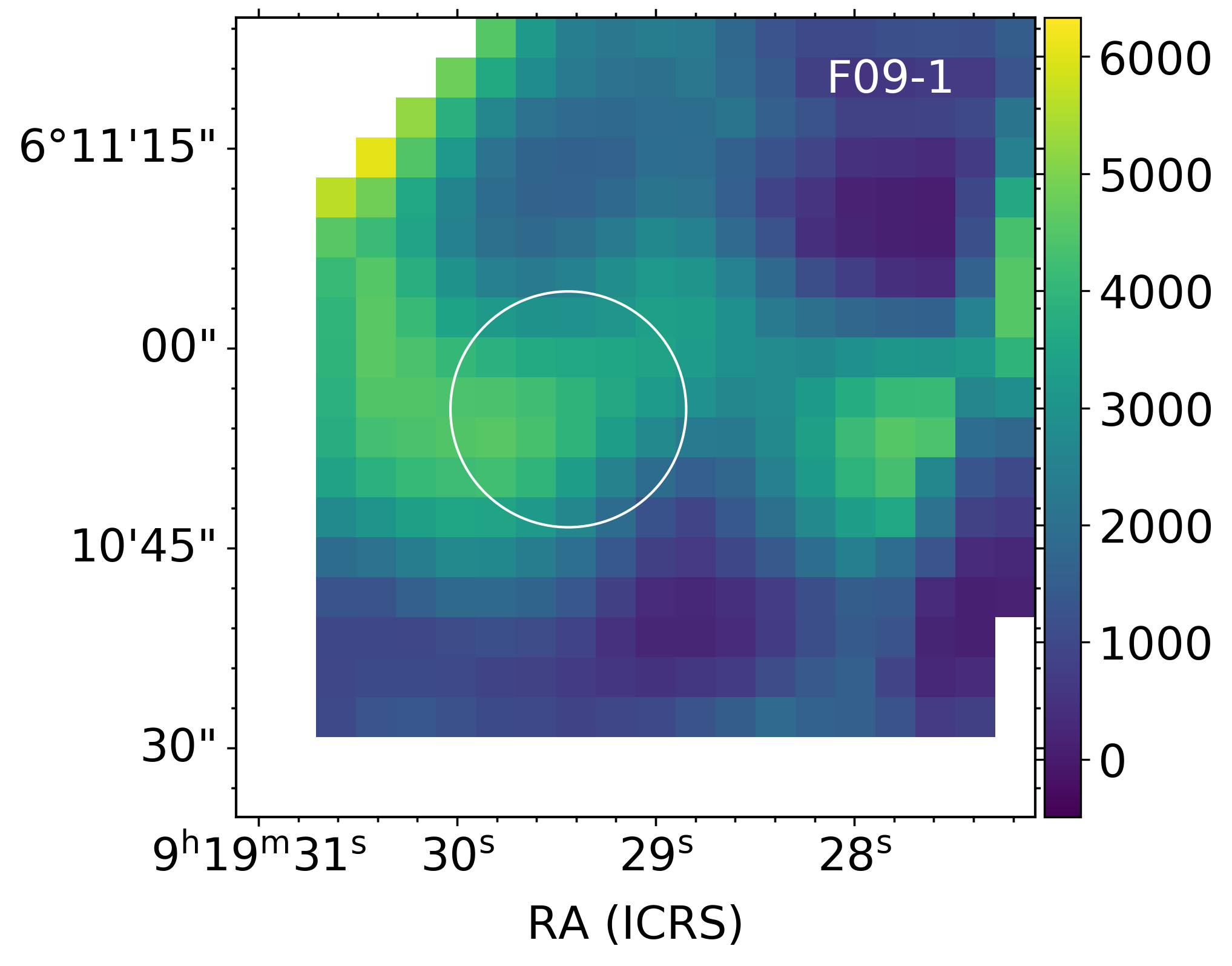}
    \includegraphics[height=4.5cm]{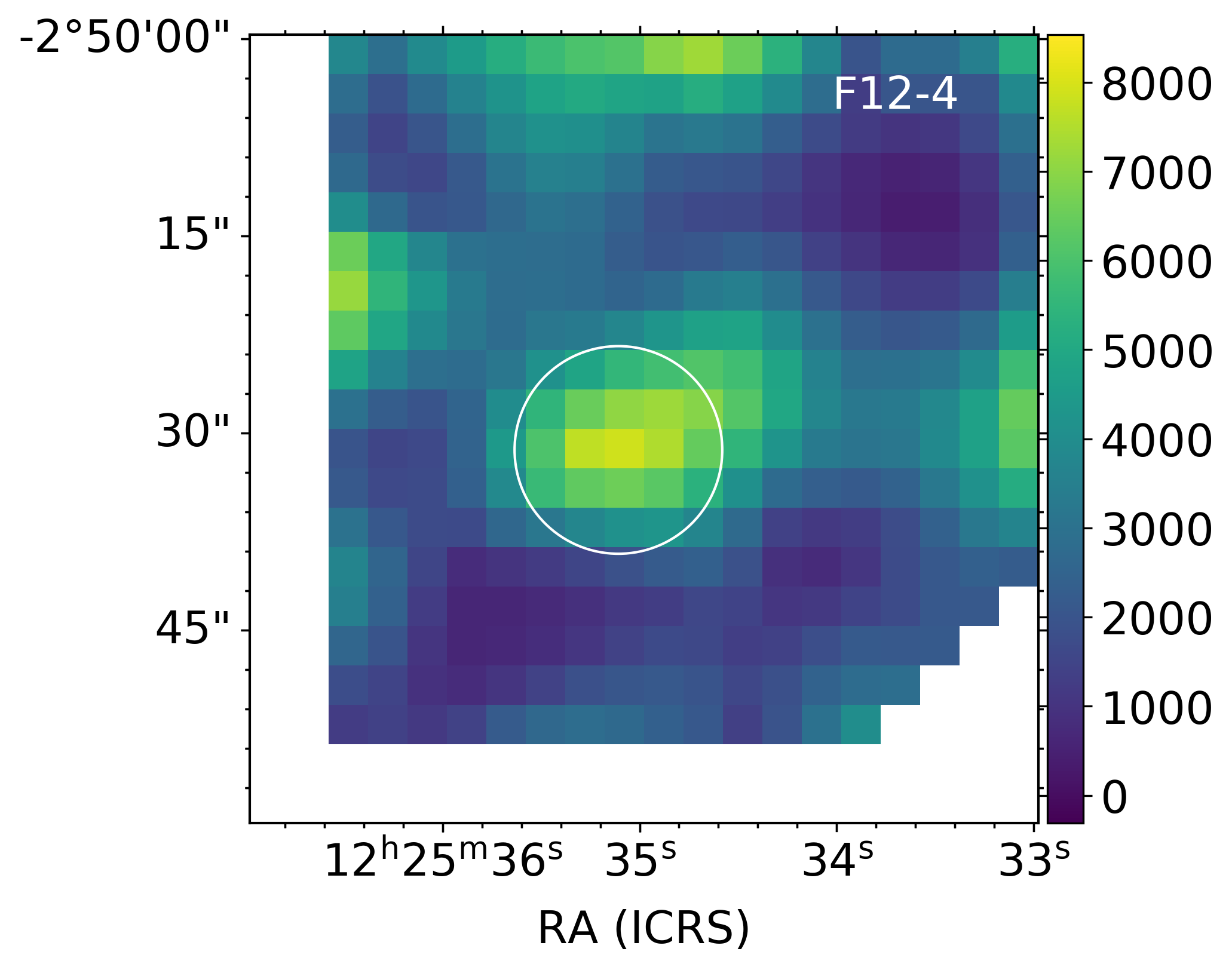}
    \caption{The \CII{} integrated intensity maps of DYNAMO galaxies in units of Jy\,km\,s$^{-1}$. The white circle is centered on the DYNAMO galaxy and its size corresponds to the angular resolution of the FIFI-LS instrument ($15.6$\arcsec{}). DYNAMO D15-3 (first row, right-most panel) is the only galaxy that overlaps with the ALMA sample.}
    \label{fig:fifi_obs}
\end{figure*}

\section{Line Ratio Literature Compilation} \label{app:lrs}
\citet{daddi15} use IRAM PdBI observations of \juptwo{}, \jupthree{}, and \jupfive{}, and Very Large Array observation of \jupone\ in three main-sequence star forming disk galaxies at $z \sim 1.5$ to study their CO excitations. We use their average $R_{31}$ and interpolate their models from their Figure 10 to extract $R_{41}$, then take the ratio $R_{41}$/$R_{31}$ to obtain $R_{43} = 0.74\, \pm\, 0.26$, which we include in Figure \ref{fig:lr_kde} as a black circle.

\citet{kamenetzky16} find a linear relation between $L_{\mathrm{FIR}}$ and $L'_{\mathrm{CO}}$ for low- to mid-J CO lines and a slightly sub-linear relation for high-J CO lines. We adopt the slope and intercepts of the relations for \jupfour{} and \jupthree{} from their Tables 6 and 7 (for U/LIRGs and non-U/LIRGs ($L_{\mathrm{FIR}} \leq 6 \times 10^{10}\,L_{\odot}$) respectively), and assume a FIR luminosity of 10$^{11}$ for the U/LIRG case and 10$^{10}$ for the non-U/LIRG case to derive the $L_{FIR}-L'_{CO}$ relations. Taking the ratio of these we find $R_{43} = 0.51\, \pm\, 0.10$ and $0.25\, \pm\, 0.05$ for U/LIRGs and non-U/LIRGs respectively, assuming 20\% uncertainty. We plot these as a black stars in Figure \ref{fig:lr_kde}.

\citet{rosenberg15} study the CO SLEDs of 29 (Ultra) Luminous Infrared Galaxies (U/LIRGs) from \jupone{} through CO(13$-$12). They classify their objects into three classes based on their excitation level. Where available, we compiled \jupfour{} and \jupthree{} fluxes from their Tables 2 and 3, and divided the resulting ratios by (J$_{u}^{3}$/J$_{l}^{3}$) to convert from units of W\,m$^{-2}$ to K. Finally, we separate the galaxies according to their classification, and plot the median line ratio for each class as black diamonds in Figure \ref{fig:lr_kde}. The error bars represent the standard deviation of line ratios in each class to illustrate the spread. We note that most of the \citet{rosenberg15} sample is contained within the \citet{kamenetzky16} sample.

\citet{papadopoulos12} study the CO SLEDs of 70 U/LIRGs; we average the $R_{43}$ values from their Table 7 (eight galaxies in total) and calculate the standard error on the mean. This results in $R_{43} = 0.96 \pm 0.12$ and plot this as a black pentagon in Figure \ref{fig:lr_kde}. We note that 11/70 galaxies from the \citet{papadopoulos12} sample overlap with the sample of \citet{rosenberg15}.

Finally, \citet{henriquez-brocal22} combine NOEMA observations of \CI{}(1$-$0), \CI{}(2$-$1), and CO(7$-$6) with ancillary \jupone{} and \jupthree{} observations to model the CO SLED of Q1700-MD94, a massive main-sequence galaxy at $z \sim 2$, with a one- and two-temperature component model using \textsc{RADEX} \citep{vanDerTak07}. We interpolate the model curves in their Figure 3 to extract $R_{43} = 0.92\, \pm\, 0.18$ and $0.77\, \pm\, 0.15$ for the one- and two-component models respectively (taking a 20\% uncertainty). We do not plot these values in Figure \ref{fig:lr_kde}, but include them in Table \ref{tab:lrs}.


\bibliography{references}{}
\bibliographystyle{aasjournal}



\end{document}